\documentclass[a4paper]{spie}  
\usepackage[]{graphicx}
\usepackage{amsmath,xspace}
\usepackage{multirow}
\usepackage{array}

\DeclareTextSymbol{\degre}{OT1}{23}
\newcommand{\mum}{\mbox{{\usefont{U}{eur}{m}{n}{\char22}}m}\xspace}
\newcommand{\imum}{\mbox{{\usefont{U}{eur}{m}{n}{\char22}}m\textsuperscript{-1}}\xspace}

\newcommand{\moy}[2][]{\left\langle#2\right\rangle_{#1}}
\newcommand{\der}[2][]{\frac{\displaystyle \mathrm d#2}{\displaystyle \mathrm
    d#1}}

\newcommand{\dd}[1]{\:\mathrm d#1}
\let\oldsqrt\sqrt
\def\sqrt{\mathpalette\DHLhksqrt}
\def\DHLhksqrt#1#2{%
\setbox0=\hbox{$#1\oldsqrt{#2\,}$}\dimen0=\ht0
\advance\dimen0-0.2\ht0
\setbox2=\hbox{\vrule height\ht0 depth -\dimen0}%
{\box0\lower0.4pt\box2}}

\title{Current results of the PERSEE testbench: the cophasing control and the
polychromatic null rate}

\author{J.~Lozi\supit{a,b,c}, F.~Cassaing\supit{b,c}, J.-M.~Le~Duigou\supit{a},
B.~Sorrente\supit{b,c}, J.~Montri\supit{b,c}, J.-M.~Reess\supit{e,c},
E.~Lhom\'e\supit{e,c}, T.~Buey\supit{e,c}, F.~H\'enault\supit{f},
A.~Marcotto\supit{f}, P.~Girard\supit{f}, M.~Barillot\supit{g},
M.~Ollivier\supit{d} and V.~Coud\'e~du~Foresto\supit{e,c}
\skiplinehalf
\supit{a}CNES - Centre National d'\'Etudes Spatiales, Toulouse, France;\\
\supit{b}Onera - The French Aerospace Lab, Ch\^atillon, France;\\
\supit{c}Groupement d'Int\'er\^et Scientifique PHASE (Partenariat Haute
r\'esolution Angulaire Sol Espace) between Onera, Observatoire de Paris, CNRS
and Universit\'e Paris Diderot;\\
\supit{d}IAS - Institut d'Astrophysique Spatiale, Orsay, France;\\
\supit{e}Observatoire de Paris - LESIA, Meudon, France;\\
\supit{f}Observatoire de la C\^ote d'Azur, Grasse, France;\\
\supit{g}Thales Alenia Space, Cannes, France
}

\authorinfo{E-mail: julien.lozi@onera.fr, Telephone: +33 (0)1 46 73 47 59}

\begin{document}
  \maketitle


\begin{abstract}
  Stabilizing a nulling interferometer at a nanometric level is the key issue
  to obtain deep null depths. The PERSEE breadboard has been designed to study
  and optimize the operation of a cophased nulling bench in the most realistic
  disturbing environment of a space mission. This presentation focuses on the
  current results of the PERSEE bench. In terms of metrology, we cophased at
  0.33~nm~rms for the piston and 80~mas~rms for the tip/tilt (0.14\% of the
  Airy disk). A Linear Quadratic Gaussian (LQG) control coupled with an
  unsupervised vibration identification allows us to maintain that level of
  correction, even with characteristic vibrations of nulling interferometry
  space missions. These performances, with an accurate design and alignment of
  the bench, currently lead to a polychromatic unpolarised null depth of
  $8.9\times10^{-6}$ stabilized at $3\times10^{-7}$ on the $[1.65-2.45]$~\mum
  spectral band (37\% bandwidth).
\end{abstract}



\section{INTRODUCTION}
\label{sec:intro}

IR spectrometry is a powerful method to characterize the hundreds of
exoplanets discovered by indirect methods, and ultimately to search for life
on telluric ones, as it allows to evidence atmospheric signatures of interest
(H$_2$O, CO$_2$, CH$_4$,...) boosted by the thermal emission of the planet.
The main challenge is to isolate the planet from its parent nearby star; one
of the very few possible methods is spaceborne nulling interferometry,
allowing implementing very efficient coronagraphs at the diffraction limit.
Within this framework, several space projects such as
DARWIN{\cite{Leger96,Leger07}}, TPF-I{\cite{Beichman99,Lawson00}},
FKSI\cite{Danchi06} or Pegase{\cite{Ollivier07,Leduigou06b}}, have been
proposed in the past years. Most of them are based on a free-flying concept to
create hectometric pupils with an acceptable launch mass. But the price to pay
is to very precisely position and stabilize the spacecrafts relatively to each
other and also with respect to an inertial direction, in order to achieve a
deep and stable extinction of the star. Understanding and mastering all these
requirements is a difficult challenge and a key issue towards the feasibility
of these missions. Thus, we decided to experimentally study this question and
focus on some possible simplifications of the concept.

Since 2006, PERSEE (Pegase Experiment for Research and Stabilization of
Extreme Extinction) laboratory test bench is under development by a consortium
composed of Centre National d'\'Etudes Spatiales (CNES), Institut
d'Astrophysique Spatiale (IAS), Observatoire de Paris-Meudon (LESIA),
Observatoire de la C\^ote d'Azur (OCA), Office National d'\'Etudes et de
Recherches A\'erospatiales (Onera), and Thales Alenia Space
(TAS){\cite{Cassaing08}}. It is mainly funded by CNES R\&D. PERSEE couples an
infrared wide band nulling interferometer, based on a Modified Mach-Zehneder
(MMZ) and a geometrical achromatic phase shifter, a disturbance simulator and
a cophasing system based on a LQG tracker for the optical path difference
(OPD). The first results of the bench were already presented in
Ref.~\citenum{Leduigou10}; in this paper, we focus on complete it with the
latest improvements and results of the bench.

After a short description of PERSEE goals and setup (focused on recent
changes), we will present in Sec.~\ref{sec:CALIBRATION} its various
calibration procedures. Recent results are reported in Sec.~\ref{sec:LQG
VALIDATION} for the cophasing system and in Sec.~\ref{sec:NULLING} for the
nuller.


\section{PERSEE DESCRIPTION AND UPDATES}
\label{sec:PERSEE DESCRIPTIONS}

This section provides the reader with a very brief description of PERSEE. Much
more details can be obtained by the interested reader from
Refs.~\citenum{Cassaing08,Houairi08b,Jacquinod08,Lozi10,Henault10}.


\subsection{PERSEE goals}
\label{sec:PERSEE goals}

The goal of PERSEE is not to reach the deepest possible nulling. Starting from
a state of the art nuller of the 2006-2007 period, it is an experimental
attempt to better master the system flowdown of the nulling requirements both
at payload (instrument main optical bench) and platform levels (satellites).
The balance of the constraints between those two levels is a key issue. The
more disturbances the payload can face, the simpler the platforms are, the
lower the cost. The general idea is hence to simplify as much as possible the
global design and reduce the costs of a possible future space mission.

The detailed objectives have been described in Ref.~\citenum{Cassaing08}. Our
main requirement is to reach a $10^{-4}$ nulling ratio in the [1.65-2.45]~\mum
band (37\% spectral bandwidth) with a $10^{-5}$ stability over a 10~h time
scale. Another important requirement is to be able to find and stabilize
fringes which have an initial drift speed (as seen from the interferometer
core) up to 150~\mum.s\textsuperscript{-1}, as this can greatly simplify the
relative metrology and control needs. We want to study and maximize the
rejection of external disturbances introduced at relevant degrees of freedom
of the optical setup by a disturbance module simulating various environments
coming from the platform level.


\subsection{Modifications of the source module}
\label{sec:modif source}

In the original design, three sources were used: a laser diode at 830~nm and a
SLED at 1320~nm for the dual band metrology system, and a black body used in
the band $[1.65-2.45]$~\mum for the science path. The three sources were
coupled in single mode optical fibers, that were joined close to each other at
the focus of the collimator, to simulate the star. That configuration was not
efficient, because the SNR in the camera was too low, the science spectrum was
polluted by the SLED flux, and the contrast was too low in the Fringe Sensor
(FS). So we had to change the configuration at the focus of the collimator, to
have only one fiber that simulates the star. Unfortunately, the black body was
not powerful enough for the FS detectors to be used for metrology in addition
to science.

The adopted solution was to use an infrared supercontinuum source, a
polychromatic source powerful enough to emit both in metrology and science
paths. Its pulse rate is at 20~MHz, much higher than the loop frequency of the
FS (1~kHz) and the camera (100~Hz). It was not possible to connectorize the
supercontinuum fiber, so we designed a simple injection bench with two off
axis parabola, to inject the flux in a single mode PCF fiber. The flux was
even too high for the science camera, so we have to defocalize the PCF fiber
to reduce the coupled light.

Another modification of the source module, presented in that
conference~\cite{Henault11}, will include a planet and a exozodical disk.


\subsection{Flux reference for the nulling rate analysis}
\label{sec:flux ref}

On the science spectrometer, two flux are measured in order to perform the
calculation of the nulling rate on each spectral channel $i$, by

\begin{equation}
  N_i = \frac{I_{min,i}}{I_{max,i}},
  \label{eq:null}
\end{equation}

where $I_{min}$ is the flux of the destructive interference, directly obtained
at the nulled output III of the MMZ. $I_{max}$ is the flux of the same output,
in the case of a constructive interference. It cannot be measured
simultaneously with $I_{min}$, so we firstly used the flux of the related
constructive interferometric output II of the MMZ. This solution has two
disadvantages: firstly, the single-mode fibers of outputs II and III must be
perfectly aligned, in order to have optimal injection in both outputs, and the
two fibers should not be subject to differential displacements. Secondly, the
output II is a coherent flux, so it is sensitive to OPD variations. Especially
in that case, the differential phase at the output II is not zero but
21\degre, thus the output II is not strictly constructive and does not measure
exactly $I_{max}$.

For those reasons, we modified that reference channel, by positioning the
reference fiber in front of the M$_0$ collimator, selecting a part of the
unused flux of the source, before the interferometric train.

The flux $I_{ref}$ of that output is not equal to the flux $I_{max}$, but they
are proportional: $I_{ref}=\gamma I_{max}$. Also, the dynamics of the camera
($\approx 10^4$) is not sufficient for the direct measurement of the desired
nulling rate, that is why we use an optical density of transmission $T$ when
we calibrate the coefficient $\gamma$, and we remove it for the measurement of
$I_{min}$. The goal is to reduce the contrast of the measurement to fit in the
range of the camera. Finally, Eq.~(\ref{eq:null}) becomes

\begin{equation}
  N_i = \frac{\gamma_i}{T_i}\frac{I_{min,i}}{I_{ref,i}}.
  \label{eq:null-frac}
\end{equation}


\section{CALIBRATION SCHEME OF THE BENCH}
\label{sec:CALIBRATION}

PERSEE bench has two control loops in piston and tip/tilt, which require a few
calibrations, as well as the scientific path. There are three types of
calibrations. The first type (Sec.~\ref{sec-spec}-\ref{sec-actu}) depends only
on the configuration, and need to be applied just once. The second type of
calibrations (Sec.~\ref{sec-tt-ref}-\ref{sec-cam}) depends slightly on the
environment, especially thermal drift. The last type of calibrations
(Sec.~\ref{sec-tt-cor}, \ref{sec-opd-ref}) is strongly affected by thermal
drift.


\subsection{Calibration of the spectral bands}
\label{sec-spec}

To calibrate the spectral bands, we used the method of Fourier Transform
Spectroscopy (FTS). The FTS uses the properties of the Fourier transform to
determine the spectrum of a source with an interferometer. In most cases, the
interferometer is a Michelson with a mirror mounted on a translation, to scan
the interference fringes. With a MMZ, the situation is slightly different. For
each wave number $\sigma$, the intensity as a function of $\delta$ is

\begin{equation}
  I(\delta,\sigma)=I(\sigma)[1+V(\sigma)\cos(2\pi\sigma\delta+\phi(\sigma))],
  \label{eq:fts1}
\end{equation}

with $I = I_a+I_b$ and $V = 2\sqrt{I_aI_b}/\left(I_a+I_b\right)$.
$I\left(\sigma\right)$ is the spectrum that we want to determine. Then the
intensity measured at the output of the interferometer is

\begin{equation}
  I(\delta)=\int_0^{\infty}I(\delta,\sigma)\dd\sigma.
\end{equation}

Then the Fourier transform of $I\left(\delta\right)$,
$\widehat{I}\left(\sigma\right)$, is

\begin{equation}
  \widehat{I}(\sigma)=I_0\delta_0(\sigma)+
  \frac12e^{i\phi(\sigma)}I(\sigma)V(\sigma),
\end{equation}

with $\delta_0$ the Dirac delta function, and $I_0 = \int_0^{\infty}
I(\sigma)\mathrm d\sigma$ the total flux. We find the case of the Michelson
interferometer, when the contrast is maximum and the phase is zero:
$\widehat{I}(\sigma) = I_0\delta_0(\sigma)+\frac12I(\sigma)$. In the case of
the MMZ, we do not measure the spectrum of the source, but the product of the
spectrum and the contrast.

\begin{figure}
  \begin{center}
    \begin{tabular}{c}
      \includegraphics[width=0.48\linewidth]{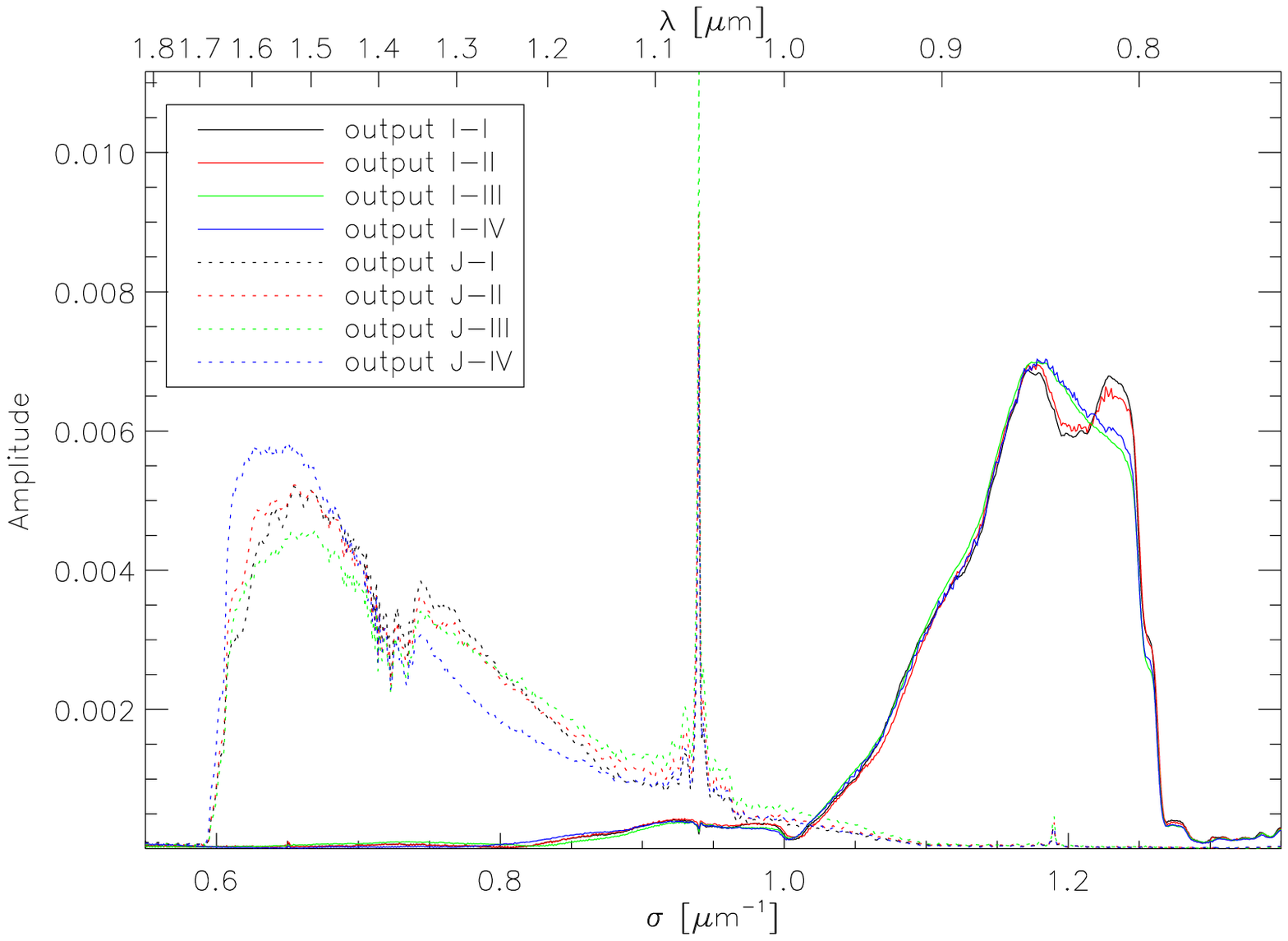}
      \includegraphics[width=0.48\linewidth]{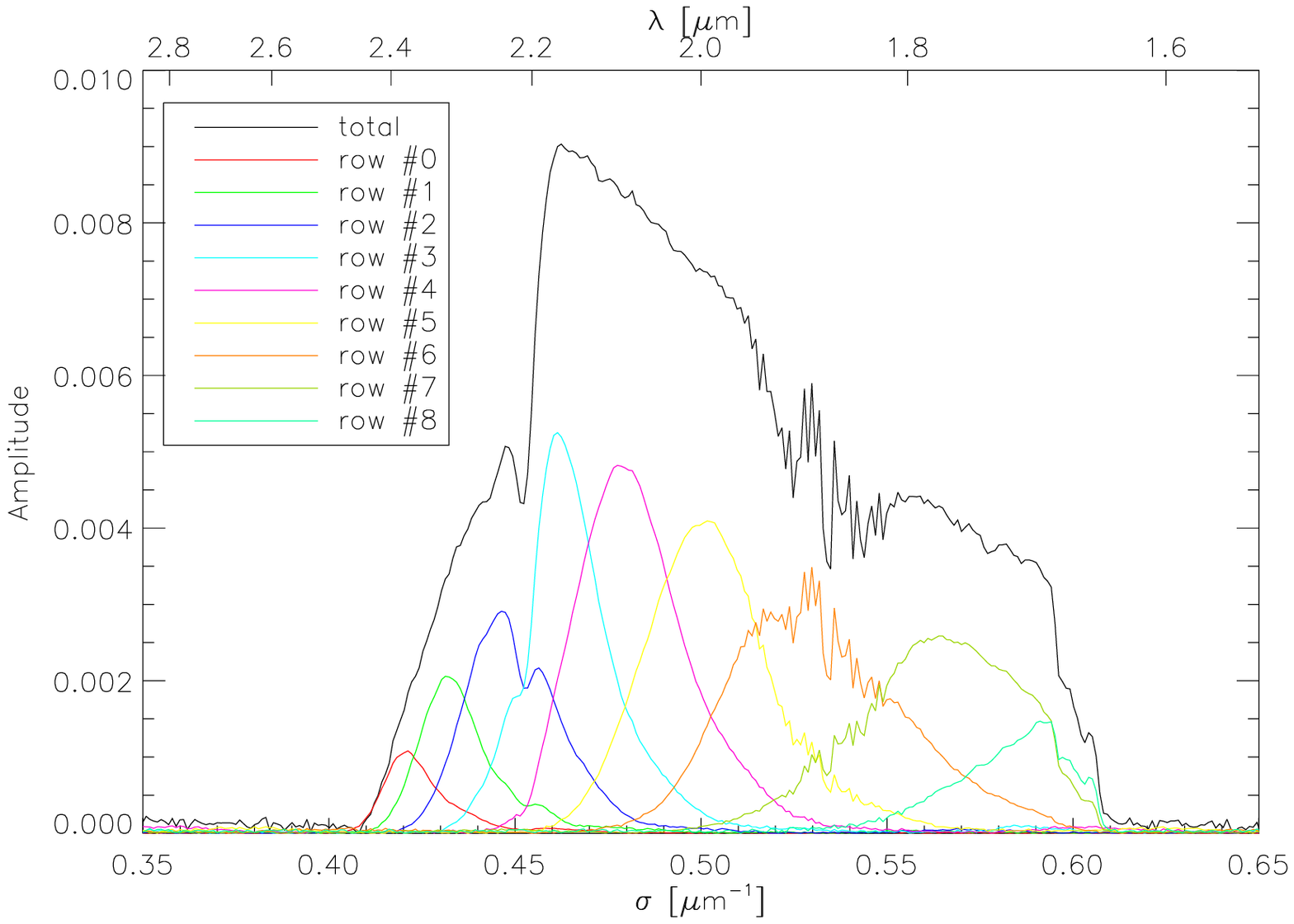}
    \end{tabular}
  \end{center}
  \caption{Left: Spectrum of the two spectral bands of the FS, which have two
    spectral bands over 4 outputs. Right: Spectra of the nine channels of the
    science camera, each one spread on two pixels on the detector.}
  \label{fig:fts}
\end{figure}

Figure~\ref{fig:fts} presents the spectra of the FS bands and the science
path. On the FS spectrum, we can see the pump laser of the supercontinuum
source, at 1064~nm, and the absorption of the water vapor around 1.4~\mum. The
four outputs of each FS spectral band are slightly different. That effect
deteriorate the linearity of the FS, especially on the J band. On the camera
spectrum, we can see that the different channels are not equal, in intensity
and bandwidth. The absorption of water vapor around 1.8~\mum is also visible.
With this analysis, we can deduce the central wavelength and the bandwidth of
each channel.


\subsection{Calibration of the density}
\label{sec-dens}

We have three optical densities on the scientific spectral band, but they were
not calibrated by the manufacturer, who did not have the appropriate
instruments in this spectral band. The theoretical absorbance of those
densities are 3, 4 and 5. So the idea is to calibrate the densities together,
to reduce the contrast between measurements. So we compared density 3 with 4,
3 with 5, and 3+4 with 5. The needed contrast is then $10^2$, which is much
less than the camera limit of $10^4$.


\subsection{Calibration of the actuators}
\label{sec-actu}

It consists in calibrating the PZT actuators of the two active M$_6$ mirrors.
Each one has three piezoelectric motors, which allow movements in piston and
tip/tilt. This calibration process measures the interaction matrix and
computes its generalized inverse to derive the command matrix. The interaction
matrix $\mathbf M_{int}$ is obtained by sending known commands (voltage ramps)
to each piezo motor, and by measuring the slope on each axis on the Field
Relative Angle Sensor (FRAS). The matrix $\mathbf M_{int}$ is then defined by

\begin{equation}
  \begin{pmatrix}
	\der[t]{x} \\[+6pt]
	\der[t]{y}
  \end{pmatrix} = \mathbf M_{int}
  \begin{pmatrix}
	\der[t]{V_1} \\[+6pt]
	\der[t]{V_2} \\[+6pt]
	\der[t]{V_3}
  \end{pmatrix}.
\end{equation}

We perform then a Singular Value Decomposition (SVD) to derive the inverse.
The piston mode is not seen by the FRAS but can be reconstructed from a
geometric model of the actuators, and its control added in the command matrix,
which transforms commands in piston and tip/tilt to voltages to be sent on the
3 piezos 120\degre apart in azimuth.


\subsection{Calibration of the tip/tilt reference position}
\label{sec-tt-ref}

The purpose of this calibration is to obtain the reference position of the
M$_6$ mirrors to inject a maximum flux in the single-mode fiber that guides
the flux from the output III of the MMZ to the scientific camera. A sequential
scan of each mirror is performed to move the focal spot in front of the fiber;
the signal is synchronously acquired and is used to reconstruct an image. The
theoretical intensity can be approximated by a gaussian
function{\cite{Ruilier01}}:

\begin{equation}
  I(\alpha,\beta) \simeq I_{inj,max}\exp\left[-\frac{\pi^2}{5}\sigma^2D^2
    \left((\alpha-\alpha_{ref})^2+(\beta-\beta_{ref})^2\right)\right],
  \label{eq-inj}
\end{equation}

with $I_{inj,max}$ the maximum injected flux, $D$ the diameter of the beam,
$\alpha$ and $\beta$ the angles of tip and tilt, and $\alpha_{ref}$ and
$\beta_{ref}$ the reference position which maximizes the injected flux.

\begin{figure}
  \begin{center}\leavevmode
 	\includegraphics[width=1.\linewidth]{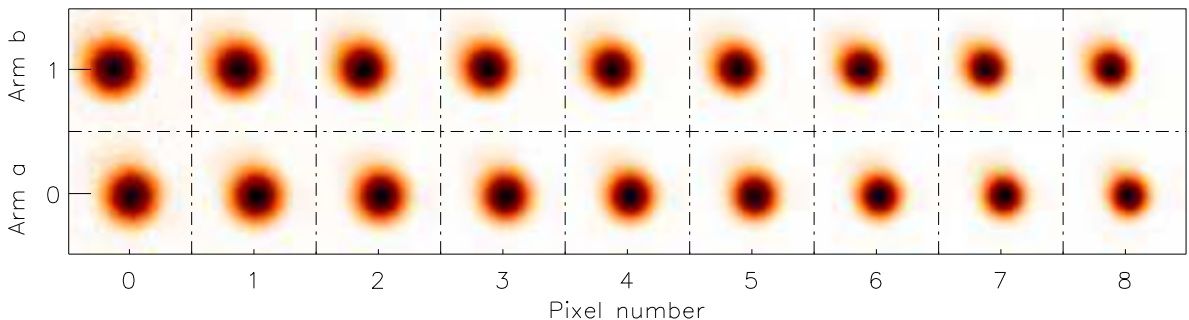}
 	\caption{Reconstruction of the convolution of the PSF and the fundamental
      guidance mode of the single-mode fiber, for each spectral channel of the
      science camera.}
 	\label{fig:recons}
  \end{center}
\end{figure}
	
The result of that reconstruction is shown in Fig.~\ref{fig:recons}, for each
channel of the science camera. We get a spot corresponding to the convolution
of the PSF for each spectral channel with the fundamental guidance mode of the
single-mode fiber. According to Eq.~(\ref{eq-inj}), the diameter of the spot
is proportionnal to $\lambda$. This can be seen in Fig.~\ref{fig:recons}: the
diameter of the spot decreases as the wavelength decreases (left to right). We
compute the centers of gravity for the two arms, which give the offsets to be
applied to both mirrors. As the spots are centro-symmetric, this position
corresponds to the maximum injection flow of each arm in the fiber. The
dispersion of the centers of gravity for the different channels are
negligible, so there is no chromatism in the reference position in tip/tilt.

	
\subsection{Calibration of the fringe sensor}
\label{sec-fs}

That procedure has already been described in Ref.~\citenum{Lozi10}. Since then
it encountered a few changes, but the idea remains the same: we inject an OPD
ramp with the M$_6$ actuators, to measure the phase shifts between the
4~outputs of the MMZ on the two spectral bands. Then we measure sequentially
the flux on each arm, and the dark current of each detector. We have then the
matrix $\mathbf M_{FS}$ given by the relation
	
\begin{equation}
  \begin{pmatrix}
    I_I\\[+2pt]
    I_{II}\\[+2pt]
    I_{III}\\[+2pt]
    I_{IV}
  \end{pmatrix}
  = \mathbf M_{FS}
  \begin{pmatrix}
    I_a\\[+2pt]
    I_b\\[+2pt]
    2\mu\mathrm{Env}(\delta)\sqrt{I_aI_b}\cos(2\pi\sigma\delta)\\[+2pt]
    2\mu\mathrm{Env}(\delta)\sqrt{I_a I_b}\sin(2\pi\sigma\delta)
  \end{pmatrix},
\end{equation}
	
with $I_a$ and $I_b$ are the flux of the two arms and
$\mu\mathrm{Env}(\delta)$ the fringes envelope. The matrix $M_{FS}$ is
inverted, to obtain the demodulation matrix D.

	
\subsection{Calibration of the science camera}
\label{sec-cam}

The calibration of the camera is made along with the calibration of the FS.
Therefore, it uses the same sequence: an OPD ramp, the flux in each arm, and
the dark current. With those information, we can analyze a few parameters, and
compute the coefficient $\gamma$ of Eq.~(\ref{eq:null-frac}) for the
calculation of the null rate.
	
With the data flux of arms $a$ and $b$, $I_a$ and $I_b$, we can estimate the
photometric unbalance $\varepsilon$ by
	
\begin{equation}
  \varepsilon = \frac{\moy[t]{I_b}-\moy[t]{I_a}}{\moy[t]{I_b}+\moy[t]{I_a}}.
\end{equation}
	
As $I_{max} = (1+V)(I_a+I_b)$, with the contrast $V\simeq1$, then $I_{max}
\simeq 2(I_a+I_b)$; we estimate the coefficient $\gamma_i$ for each spectral
channel by
	
\begin{equation}
  \gamma_i =
  \frac{\moy[t]{I_{ref,i}}}{2\left(\moy[t]{I_{a,i}}+\moy[t]{I_{b,i}}\right)}.
\end{equation}
	
\begin{figure}
  \begin{center}\leavevmode
 	\includegraphics[width=.5\linewidth]{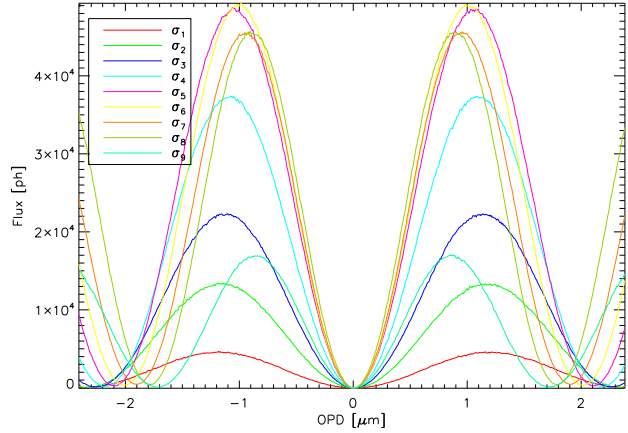}
 	\caption{Interferogram acquired during the calibration of the camera.}
 	\label{fig:interf-III}
  \end{center}
\end{figure}
	
Figure~\ref{fig:interf-III} presents an interferogram obtained on the camera
during the OPD ramp of the calibration. We can see that every channel has
different wavelengths, but their minima are at the same position. With those
data, we can use a FTS mode, to determine the central frequency $f_i$ and the
phase $\phi_i$ of each sinusoid.
	
With the frequency $f_i$, we can determine the wave number $\sigma_i$ with the
slope of the OPD ramp $v$: $\sigma_i = f_i/v$. With the phase $\phi_i$, we can
determine the position of the minimum of each sinusoid, by
	
\begin{equation}
  \delta_i = \delta_{ref}+\frac{\phi_i}{2\pi\sigma_i},
  \label{eq-delta-i}
\end{equation}
	
with $\delta_{ref}$ is the reference position of the FS.


\subsection{Correction of the tip/tilt reference position}
\label{sec-tt-cor}

When the temperature drifts, the reference position of the FRAS also drifts.
This shift is usually small, so it is not necessary to repeat the procedure
described above. A more efficient procedure has been developed for this
purpose. It is inspired by the procedure of correction of the OPD position
described below, called dithering{\cite{Gabor08}}.
	
Close to the reference position, we can approximate the exponential term of
Eq.~(\ref{eq-inj}) with a quadratic term:
	
\begin{equation}
  I(\alpha,\beta) \simeq I_{inj,max}\left(1-\frac{\pi^2}{5}\sigma^2D^2
    \left[(\alpha-\alpha_{ref})^2+(\beta-\beta_{ref})^2\right]\right)
  \label{eq:flux-angle}
\end{equation}
	
For the two arms, for both axis, we put a small angle $\theta=6$~arcsec (11\%
of the Airy disk), and we measure the flux injected in the fiber. If
$\alpha_0$ and $\beta_0$ are the current positions, we note
$I_0=I(\alpha_0,\beta_0)$, $I_{\alpha_-} = I(\alpha_0-\theta,\beta_0)$,
$I_{\alpha_+} = I(\alpha_0+\theta,\beta_0)$, $I_{\beta_-} =
I(\alpha_0,\beta_0-\theta)$ and $I_{\beta_+} = I(\alpha_0,\beta_0+\theta)$.
	
Then, we calculate the reference position with
	
\begin{equation}
  \left\{
    \begin{array}{rcl}
      \alpha_{ref} &=& \alpha_0-\frac{\displaystyle I_{\alpha_+}-I_{\alpha_-}}
      {\displaystyle 2\left(I_{\alpha_+}+I_{\alpha_-}-2I_0\right)}\cdot\theta
      \\[+12pt]
      \beta_{ref} &=& \beta_0-\frac{\displaystyle I_{\beta_+}-I_{\beta_-}}
      {\displaystyle 2\left(I_{\beta_+}+I_{\beta_-}-2I_0\right)}\cdot\theta \\
    \end{array}
  \right..
\end{equation}

This procedure was successfully implemented on the bench, and showed a very
good accuracy ($\approx 50$~mas, 0.1\% of the Airy disk).

	
\subsection{Correction of the OPD reference position}
\label{sec-opd-ref}

Despite the calibration procedure of the camera described above, the accuracy
is not enough to guaranty a sub-nanometer positioning of the OPD to minimize
the null rate. That position is also very sensitive to temperature variations.
To correct the reference position of the FS $\delta_{ref}$, we use the
technique of dithering{\cite{Gabor08}}.
	
The null rate has a quadratic dependence with the OPD{\cite{Lawson00}}:

\begin{equation}
  N(\delta)=N_{min}+(\pi\sigma(\delta-\delta_{ref}))^2,
\end{equation}
	
So we measure the null rate for three positions of OPD: the current position
$\delta_0$, and two other positions $\delta_0-\varepsilon$ and
$\delta_0+\varepsilon$, with $\varepsilon=2$~nm. If we note $N_0 =
N(\delta_0)$, $N_{-\varepsilon} = N(\delta_0-\varepsilon)$ and
$N_{+\varepsilon} = N(\delta_0+\varepsilon)$, the reference position is then
	
\begin{equation}
  \delta_{ref} = \delta_0-\frac{N_{+\varepsilon}-N_{-\varepsilon}}
  {2\left(N_{+\varepsilon}+N_{-\varepsilon}-2N_0\right)}\cdot\varepsilon
\end{equation}

That reference position is very sensitive to temperature variations, about
600~nm.K\textsuperscript{-1}. However, we consider that temperature variations
over 100~s are less than 1~mK in normal conditions, so the thermal drift is
less than 0.6~nm over 100~s.


\section{LQG EXPERIMENTAL VALIDATION}
\label{sec:LQG VALIDATION}

PERSEE aims to simulate a nulling space mission, constituted of three
satellites: two siderostats and a central combiner. The three satellites are
controlled by impulsive gas thrusters and reaction wheels, from individual
star trackers, a radiofrequency metrology, optical lateral metrology and
off-loading signals from the central combiner. Those systems lead to different
types of disturbances, viz., microvibrations due to the reaction wheels,
linear-parabolic drift due to the solar pressure and the gas impulses, and
low-frequency pointing control (PC) residue.

Vibrations are the main disrupters of the OPD, which are not always well
corrected by an integrator in a control loop. We implemented the method
described in Ref.~\citenum{Lozi10}, based on the work for the extreme adaptive
optics instrument SPHERE at Onera{\cite{Meimon10}}. It is based on a Linear
Quadratic Gaussian (LQG) control law that provides an optimal correction of
vibrations, in the sense of residual phase minimum variance.


\subsection{Identification of the pointing control residue and vibration
filtering efficiency}
\label{sec-identification-pointing}

In a first test, we analyze the correction of the PC residue, with a classical
integrator and a LQG.

We note $\sigma_{OL,dis}^2$ the open-loop OPD variance for a case where we
inject a PC residue of variance $\sigma_{inj,PC}^2$, and vibrations of
variance $\sigma_{inj,vib}^2$. If $\sigma_{OL,exp}^2$ is the open-loop OPD
variance of the experiment without any injected disturbance, then

\begin{equation}
  \sigma_{OL,dis}^2 = \sigma_{OL,exp}^2+\sigma_{inj,PC}^2+\sigma_{inj,vib}^2.
  \label{eq-sig-dis}
\end{equation}

So if $\sigma_{OL,PC}^2$ is the open-loop OPD variance for the case where we
inject only PC residue, we have $\sigma_{OL,PC}^2 =
\sigma_{OL,exp}^2+\sigma_{inj,PC}^2$, and then $\sigma_{OL,dis}^2 =
\sigma_{OL,PC}^2+\sigma_{inj,vib}^2$.

For this test, we compute the following power residue percentage

\begin{equation}
  \rho_{c,PC} = \frac{\sigma_{c,PC}^2-\sigma_{c,exp}^2}
  {\sigma_{OL,PC}^2-\sigma_{c,exp}^2}, \ c \in \{int, LQG\},
  \label{eq-rho-PC-c}
\end{equation}

with $\sigma_{c,exp}^2$ the residual OPD variance for the controller $c$
(integrator or LQG) when we do not inject disturbances, and $\sigma_{c,PC}^2$
the same quantity for the case where we inject the PC residue. In that case,
if $\sigma_{c,PC}^2 = \sigma_{c,exp}^2$, then $\rho_{c,PC} = 0$\%, the
injected disturbance is totaly corrected, and if $\sigma_{c,PC}^2 =
\sigma_{OL,PC}^2$, then $\rho_{c,PC} = 100$\%, the controller has no effect.

\begin{table}
  \begin{center}
    \begin{tabular}{ccc}
      \multicolumn{3}{c}{POINTING CONTROL RESIDUE CORRECTION}\\
      \hline
      Controller $c$ & $\sigma_{c,PC}$ & $\rho_{c,PC}$ \\
      \hline
      Integrator & 0.45~nm~rms & 0.17\% \\
      LQG & 0.37~nm~rms & 0.09\% \\
    \end{tabular}
    \caption{Correction of the PC residue with the integrator and the LQG.}
    \label{tab:cor-PC}
  \end{center}
\end{table}

Numerical results of the two controllers are presented in
Tab.~\ref{tab:cor-PC}. With that type of disturbance, i.e. with a continuously
decreasing spectrum like in AO with turbulence, the integrator and the LQG
have similar performance: the power residue is very low in each case, with a
slight improvement from the LQG.

For the following tests, we will compute a vibration residue percentage,
similar to the one used in Ref.~\citenum{Meimon10}:

\begin{equation}
  \rho_c = \frac{\sigma_{c,dis}^2-\sigma_{c,PC}^2}
  {\sigma_{OL,dis}^2-\sigma_{OL,PC}^2-\sigma_{c,PC}^2}, \ c \in \{int, LQG\}
  \label{eq-rho-c}
\end{equation}

with $\sigma_{c,dis}^2$ is the residual OPD variance for the controller $c$
for the case where we inject the PC residue with vibrations.So if
$\sigma_{c,dis}^2 = \sigma_{c,PC}^2$, the $\rho_i = 0$\%, the vibrations are
totally removed. On the contrary, if $\sigma_{c,dis}^2 = \sigma_{inj,vib}^2$,
$\rho_c = 100$\%, the vibrations are not corrected.


\subsection{Experimental correction of a single vibration}
\label{sec-single-vib}

In those tests, we inject the PC residue with one vibration, at several
frequencies, and we compare the correction of the integrator and the LQG.

\begin{table}
  \centering
  \begin{tabular}{ccccc}
    \multicolumn{5}{c}{CORRECTION OF A SINGLE VIBRATION}\\
    \hline
    $f_{vib}$ & $\sigma_{int,dis}$ & $\rho_{int}$ & $\sigma_{LQG,dis}$ &
    $\rho_{LQG}$ \\\relax
    [Hz] & [nm rms] & [\%] & [nm rms] & [\%] \\
    \hline
    1.1 & 0.49 & 0.08 & 0.38 & 0.02 \\
    2.2 & 0.58 & 0.34 & 0.40 & 0.06 \\
    5.5 & 0.87 & 1.2 & 0.47 & 0.17 \\
    11 & 1.5 & 8.9 & 0.38 & 0.03 \\
    22 & 3.0 & 17 & 0.52 & 0.27 \\
    55 & 7.1 & 92 & 0.65 & 0.53 \\
    110 & 11 & 220 & 0.4 & 0.04 \\
    220 & 4.4 & 130 & 0.38 & 0.04 \\
  \end{tabular}
  \caption{Correction of single vibrations at different frequencies.}
  \label{tab:freq-vib}
\end{table}

Table~\ref{tab:freq-vib} presents the results of the integrator and the LQG
for different vibration frequencies. As expected, the correction of the
integrator deteriorates as the frequency increases, in contrast to LQG, which
maintains a fairly constant performance $\rho_{LQG}$ close to 0\%.

For $f_{vib} = 1.1$ and 2.2~Hz, the injected vibration was not identified, but
corrected as a part of the PC residue contribution (the continuum part); it
does not affect the result of the correction: the disturbance is totally
removed.

Table~\ref{tab:freq-vib} shows that after 22~Hz, for 10~nm vibrations, the
integrator is not efficient to correct the disturbances at the required level
of 1.8~nm~rms. Also, the power of the vibration at 110~Hz is doubled, because
of the overshoot in the integrator closed-loop transfer function.

These tests revealed a limitation of identification: the spectral resolution
of the FFT. Indeed, if a sinusoid has not an integer number of periods in the
signal supplied to the identification, the fitted vibration will have a
damping factor greater than reality.

\begin{figure}
  \begin{center}
    \begin{tabular}{c}
      \includegraphics[width=0.48\linewidth]{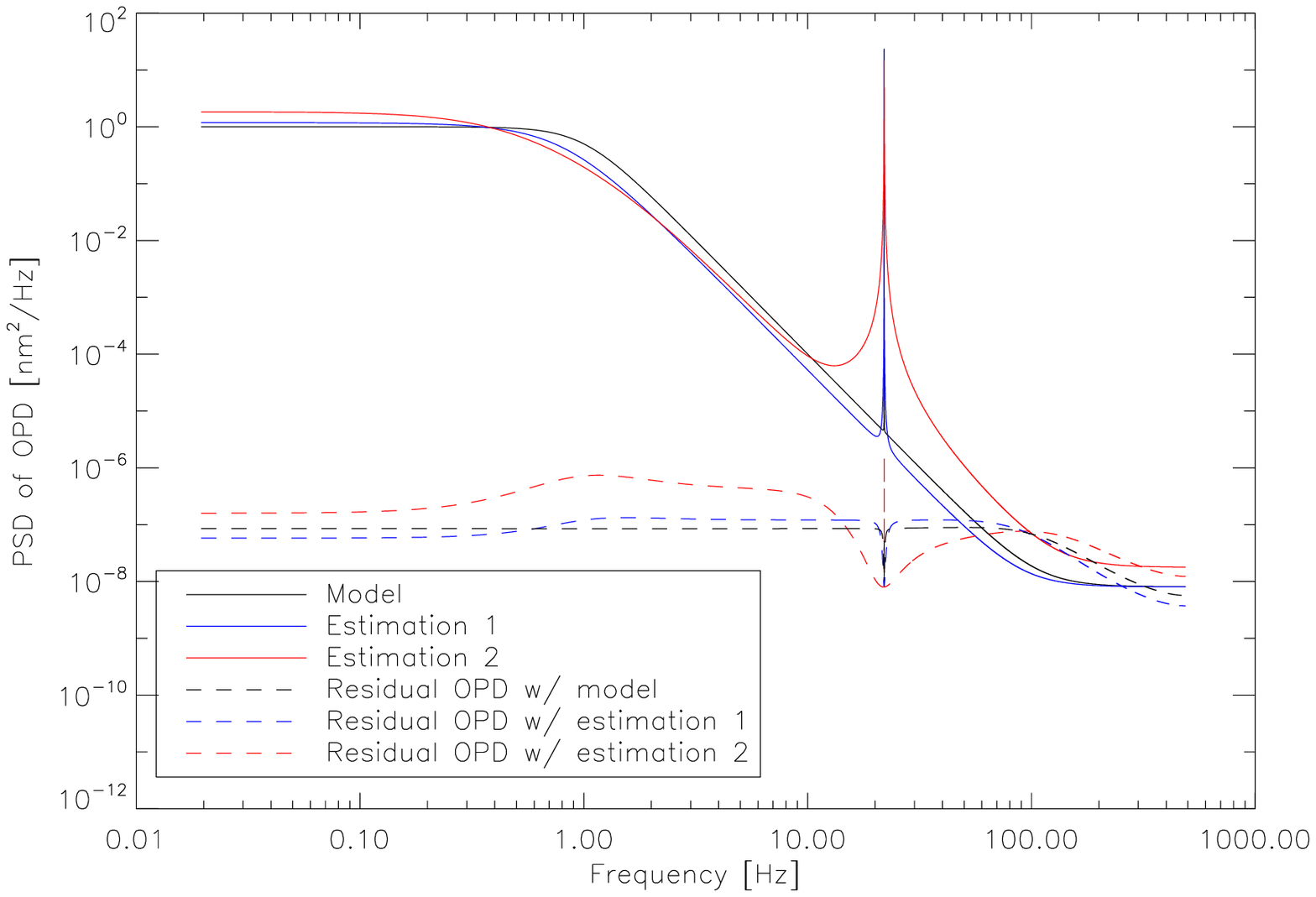}
      \includegraphics[width=0.48\linewidth]{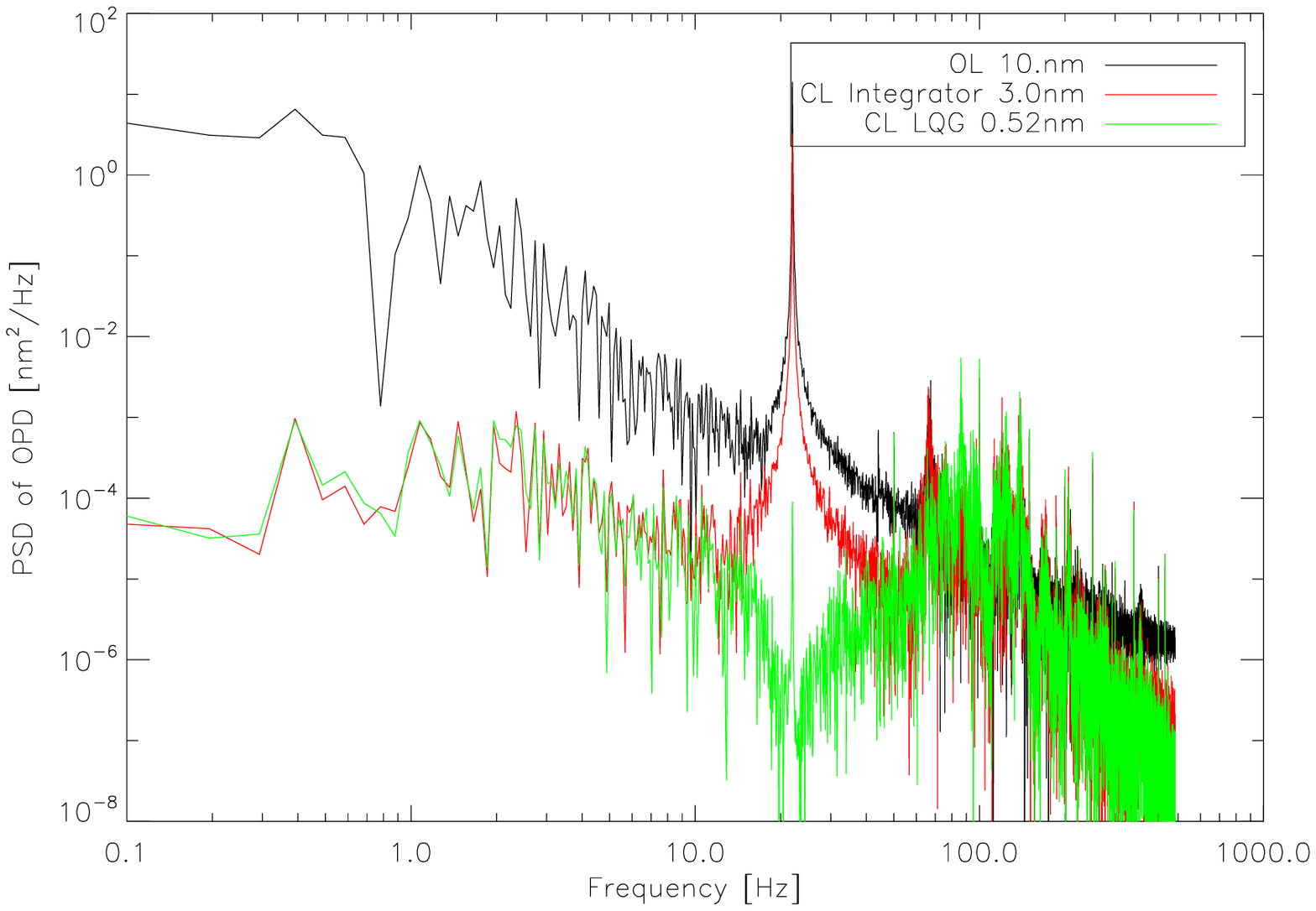}
    \end{tabular}
  \end{center}
  \caption{Frequency evolution of the identification, and on the residual OPD
    (simulation and experimental results).}
  \label{fig-p-sim-kvib}
\end{figure}

Figure~\ref{fig-p-sim-kvib} illustrates that effect. In that example, the
model used is a typical PC residue, with a single vibration around 22~Hz
(black solid line). For the estimation~1 (blue solid line), the vibration has
an integer number of periods in the time sequence used for identification. The
estimation is then close to the model for the PC residue, and the identified
damping coefficient for the vibration is $10^{-5}$. Then the OPD residue is
close to the ideal one, giving almost a white noise. For the estimation~2 (red
solid line), the number of periods is not aninteger, the peak maximum is
between two points of the periodogram. In that case, because of the contrast
between the vibration and the PC residue, the peak is not sharp, but has some
leakages around the central frequency. The identification is degraded, the PC
residue is not perfectly fitted, especially at the cutoff frequency. For the
vibration, the identification gives a greater damping coefficient of
$5\times10^{-4}$, and a reduced amplitude. On the OPD residue, we can see that
the signal is overcorrected around the vibration frequency, while the
vibration is not perfectly suppressed. This overcorrection is compensated by a
higher residue at low frequencies.


\subsection{Experimental correction of realistic disturbances}
\label{sec-simulation-results}

These tests aim to simulate as accurately as possible the conditions of
disturbance of a space mission. To do so, we consider two levels of PC
residue: an optimistic value of 3~nm~rms, and a conservative value of
15~nm~rms.

Added to that residue, we also inject the characteristic vibrations of 6
reaction wheels, rotating at different frequencies, close to a central
frequency $f_{rw}$. We consider three rotation frequencies, 1.5, 3 and 4.5~Hz,
corresponding to typical GNC design values. Those reaction wheels generate a
set of vibrations, related to the first harmonic of the wheel.

No detailed mechanical model was available, and the following assumptions were
made from typical satellite behavior known at CNES. Some of the frequencies
are amplified by structure modes in the satellites: a mode of sunshade at
15~Hz, a fundamental lateral mode of the satellite at 55~Hz, and local modes
of the wheel cavity and the optics at 75 and 85~Hz.

Those parameters lead to six different cases that are studied with a simple
simulator of the bench, and compared to experimental results. For a central
frequency $f_{rw} = 1.5$~Hz, only the first harmonics are significant, no
harmonic is amplified by the structure modes. For $f_{rw} = 3$~Hz, some
harmonics are amplified at 55~Hz, while for $f_{rw} = 4.5$~Hz, harmonics are
also amplified at 75 and 85~Hz.

\begin{figure}
  \begin{center}
    \begin{tabular}{c}
      \includegraphics[width=0.48\linewidth]{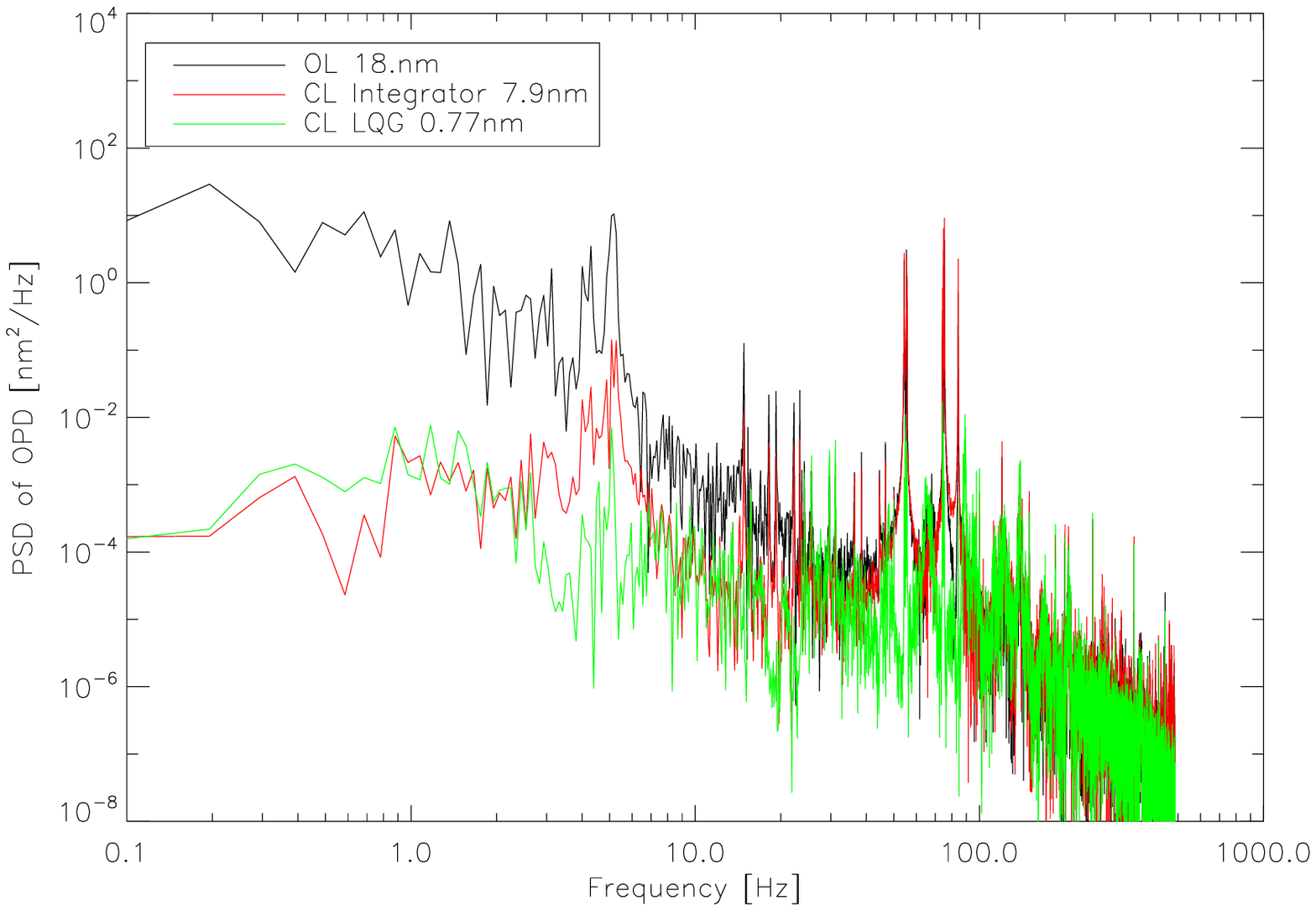}
      \includegraphics[width=0.48\linewidth]{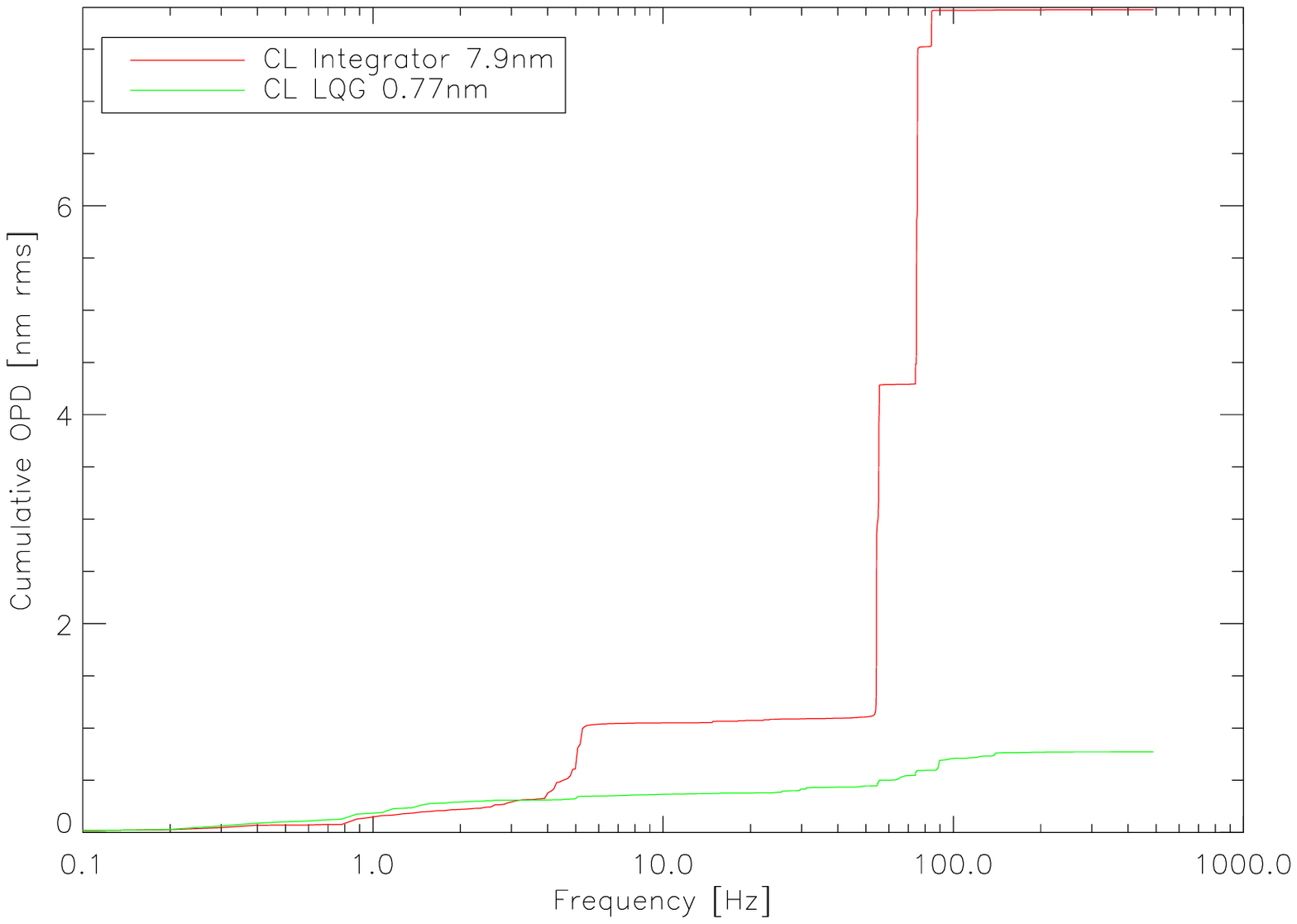}
    \end{tabular}
  \end{center}
  \caption{PSD and cumulative OPD for the correction of realistic disturbance,
    with an integrator and a LQG. The parameters of the disturbance are
    $f_{rw} = 4.5$~Hz and $\sigma_{OL,PC} = 15$~nm~rms.}
  \label{fig-pi-oil-microvib}
\end{figure}

Figure~\ref{fig-pi-oil-microvib} presents the correction of the worst studied
case: $f_{rw}=4.5$~Hz and $\sigma_{OL,PC}=15$~nm~rms. For the LQG, the
identification is made on the 20 strongest vibrations. We set the integrator
gain at 0.5.

\begin{table}
  \centering
  \begin{tabular}{cccccc}
    \multicolumn{6}{c}{COMPARISON B/W SIMULATIONS AND EXPERIMENTS}\\
    \hline
    $f_{rw}$ & $\sigma_{OL,PC}$ & $\sigma_{int,dis}$ & $\rho_{int}$ &
    $\sigma_{LQG,dis}$ & $\rho_{LQG}$ \\\relax
    [Hz] & [nm~rms] & [nm~rms] & [\%] & [nm~rms] & [\%] \\
    \hline
    1.5 & 3 & 0.62 & 0.54 & 0.41 & 0.09 \\
    1.5 & 15 & 0.58 & 0.06 & 0.44 & 0.03 \\
    3.0 & 3 & 2.2 & 11 & 0.47 & 0.20 \\
    3.0 & 15 & 4.8 & 13 & 0.51 & 0.07 \\
    4.5 & 3 & 4.3 & 58 & 0.49 & 0.34 \\
    4.5 & 15 & 7.9 & 22 & 0.77 & 0.17 \\
  \end{tabular}
  \caption{Influence of the number of vibrations.}
  \label{tab:real-dist}
\end{table}

Table~\ref{tab:real-dist} presents the comparison between the simulations and
the experimental results for the 6 different cases. We can see that the
integrator can manage the disturbance at the required levels for $f_{rw} =
1.5$~Hz, whatever the level of RF PC residue. For those cases, the LQG gives a
substantial improvement, leading to a nulling contribution divided by 2. The
interest of LQG is more significant for the other cases, because amplified
harmonics are not properly corrected by the integrator. As expected, the LQG
corrector removes almost the whole power of the disturbance.


\section{NULLING ANALYSIS}
\label{sec:NULLING}


\subsection{Nulling calculation and contributors}
\label{sec:calculation}

On PERSEE, we carry out a spectral analysis of the null rate, so we analyze
separately the different spectral channels. Thus, we calculate an average null
rate over the entire band, rather than an integrated null rate, which is more
dependant of the spectrum. Therefore, the spectral bandwidth is larger.

With the result of Fig.~\ref{fig:fts}, we calculate the spectral widths
$\left(\frac{\Delta\sigma}{\sigma}\right)_i$ for each channel $i$. Then the
total spectral bandwidth is deduced from the extreme channels:

\begin{equation}
  \Delta\sigma_m =
  \sigma_n+\frac{\Delta\sigma_n}{2}-\sigma_1+\frac{\Delta\sigma_1}{2},
\end{equation}

with $n$ the number of channel ($n=9$). Then, the total bandwidth is
$\left(\frac{\Delta\sigma}{\sigma}\right)_m = 37$\% around the central wave
number $\sigma_m = 0.5$~\imum, i.e. a central wavelength $\lambda_m =
2.0$~\mum. For the analysis of the bench performance, we will then compute the
mean null rate $N_m = \moy[i]{N_i}$.

The null rate can be decomposed in a sum of independent
contributors{\cite{Serabyn01}}:

\begin{equation}
  N_m(t) = \frac14 \left[ \Delta\phi_c(t)^2 +
    \moy[t]{\Delta\phi_\lambda(t)^2}+\frac{\pi^2}{4}\left(\theta_{dia}
      \sigma_{max}b\right)^2+\frac{1}{4}(\Delta\phi_{s-p})^2+\alpha^2_{rot}
    +\varepsilon(t)^2\right],
  \label{eq-system-nulling}
\end{equation}

where:

\begin{itemize}
\item $\Delta\phi_c(t)$ is the phase fluctuation at the central wavelength,
  while $\Delta\phi_\lambda(t)$ takes into account the chromatic dispersion of
  the null;

\item $\theta_{dia}$ is the angular diameter of the source, $\sigma_{max}$ the
  maximum wavenumber of the spectrum and $b$ the interferometric baseline;

\item $\Delta\phi_{s-p}$ is the phase difference between the two polarizations
  $s$ et $p$, and $\alpha_{rot}$ is the rotation angle of the polarization $s$
  (or $p$) between the two beams of the interferometer;

\item $\varepsilon(t)$ are the fluctuations of intensity mismatch, defined by
  $\varepsilon = \frac{I_b-I_a}{I_b+I_a}$, with $I_a$ and $I_b$ the flux on
  the arms $a$ and $b$.
\end{itemize}

\begin{figure}[b]
  \begin{center}
    \begin{tabular}{c}
      \includegraphics[width=0.48\linewidth]{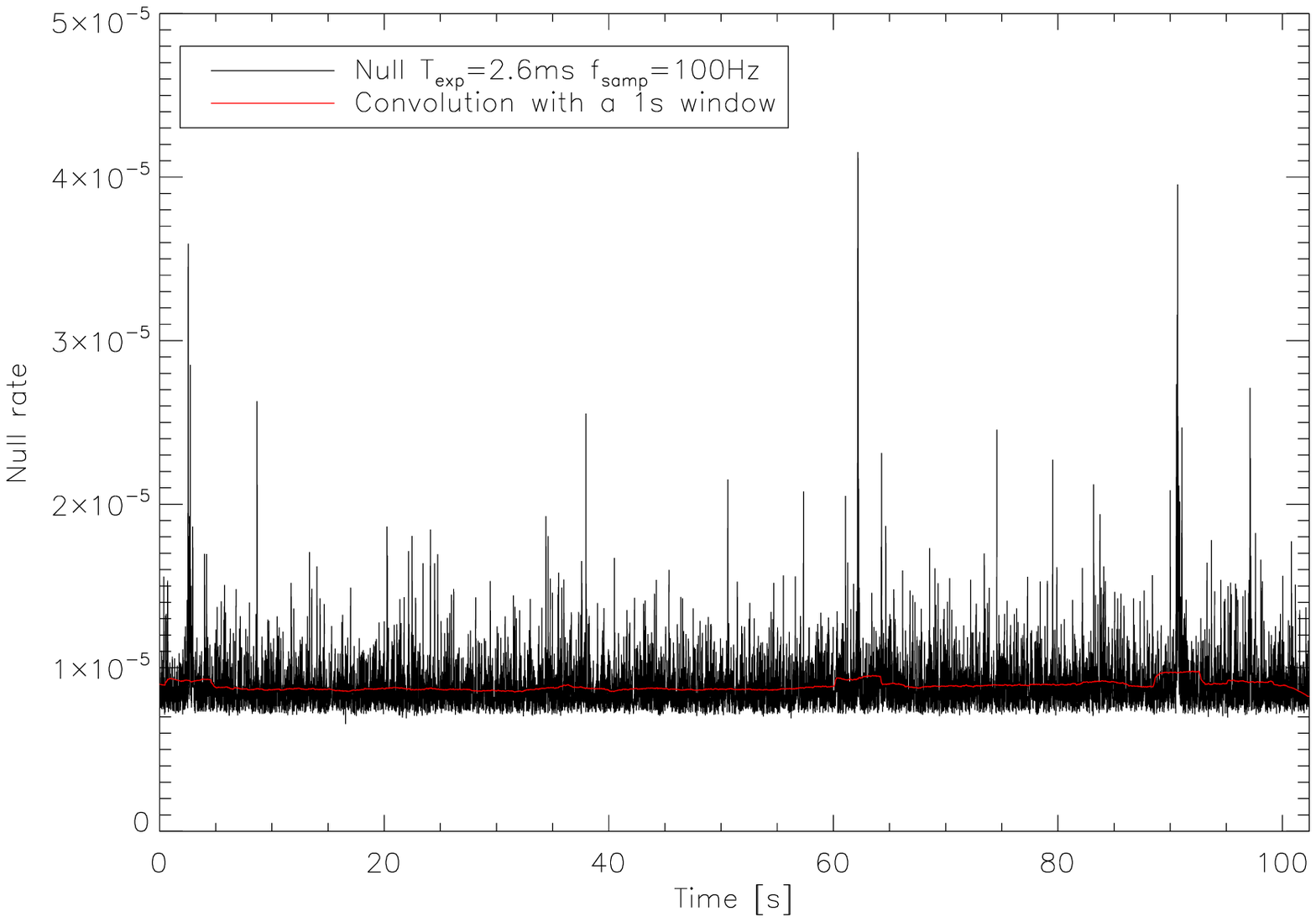}
      \includegraphics[width=0.48\linewidth]{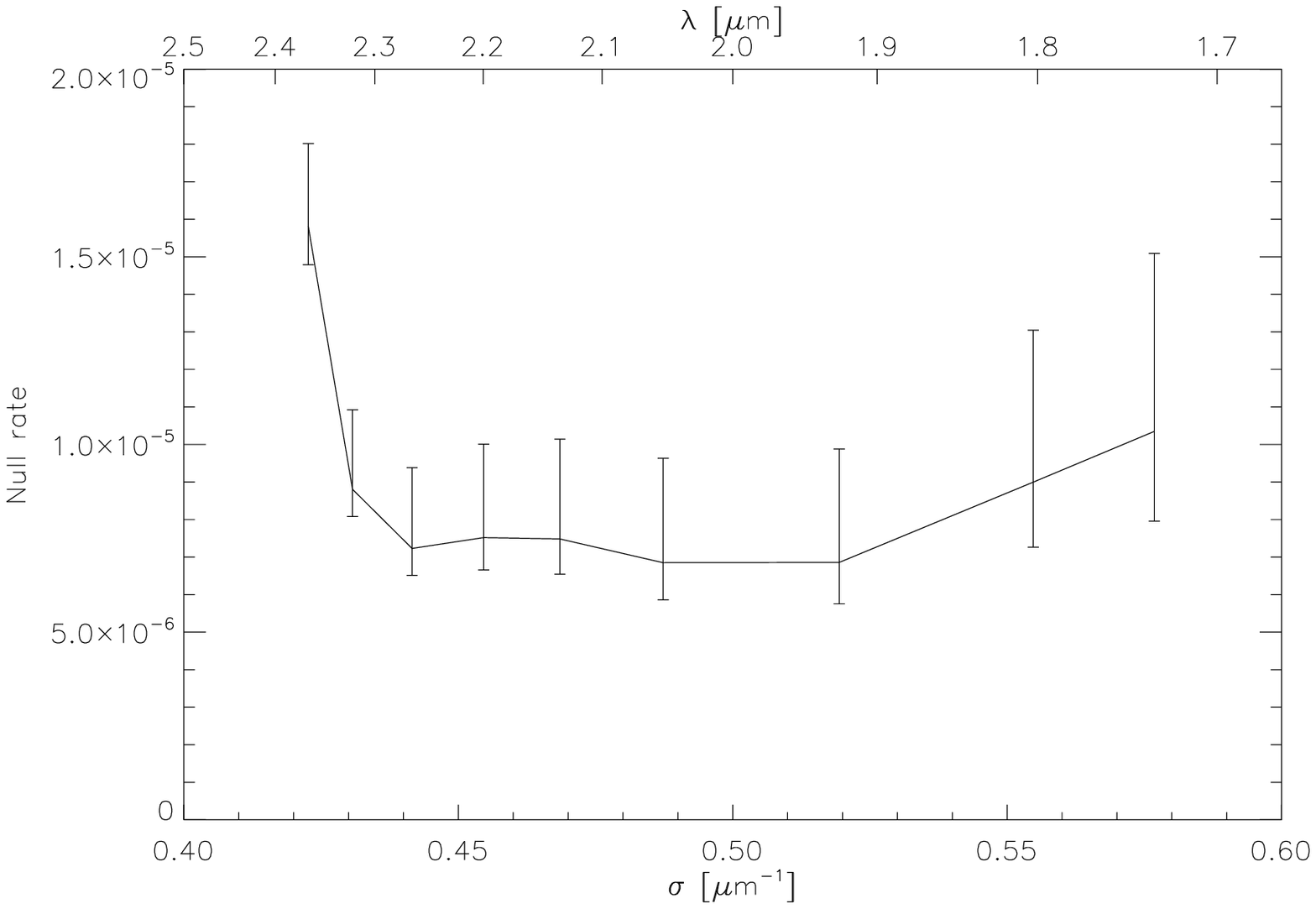}
    \end{tabular}
  \end{center}
  \caption{Best result of null rate on 100~s: temporal sequence of the mean
    null rate (left) and temporal average of the null rates on the different
    spectral channels (right).}
  \label{fig-best-null}
\end{figure}

Figure~\ref{fig-best-null} presents the best result obtained on the bench, on
a 100~s timescale, with an acquisition frequency of 100~Hz, and an integration
time of 2.58~ms. The mean value is

\begin{equation*}
  \moy[t=100s]{N_m} = 8.9\times10^{-6}.
\end{equation*}

The standard deviation $\sigma_{N,T}$ depends on the integration time $T$. We
chose to consider $T=1$~s, more representative of a real mission. Then, we
have $\sigma_{N,T=2.58\mathrm{ms}}=1.8\times10^{-6}$ and
$\sigma_{N,T=1\mathrm{s}}=2.7\times10^{-7}$. The chromatic dispersion of the
null rate is between $6.9 \times 10^{-6}$ and $1.58 \times 10^{-5}$ on the
spectrum. the degradation of the null rate at short wavelengths is explained
by both OPD and flux mismatch effects, but the degradation of the highest
wavelength is probably due to polarization effects.


\subsection{OPD fluctuations and chromatism}

The OPD fluctuations are an important contribution of the null rate. The
different spectral channels are disturbed differently by the OPD fluctuations.

\begin{figure}[b]
  \begin{center}
    \begin{tabular}{c}
      \includegraphics[width=0.48\linewidth]{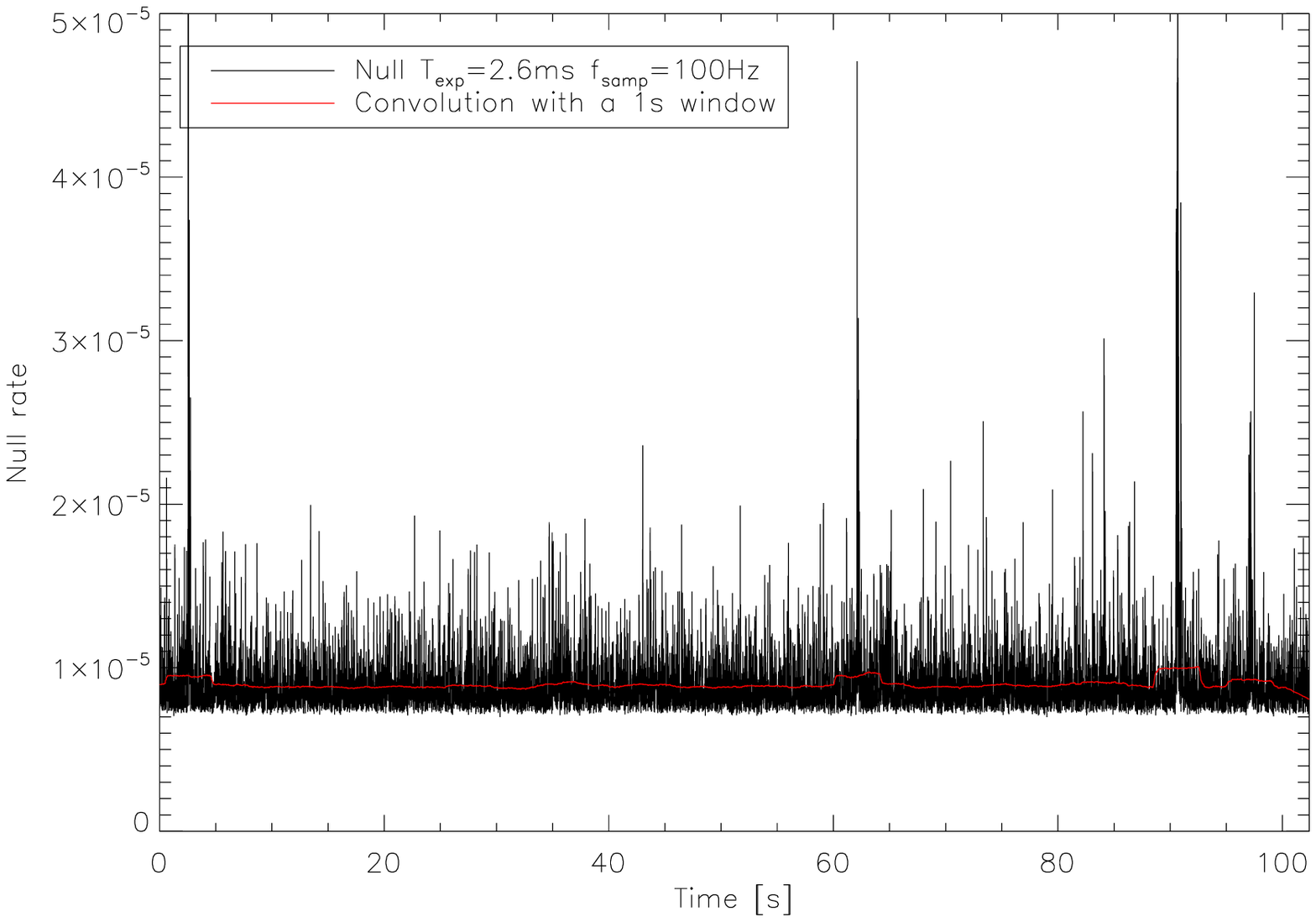}
      \includegraphics[width=0.48\linewidth]{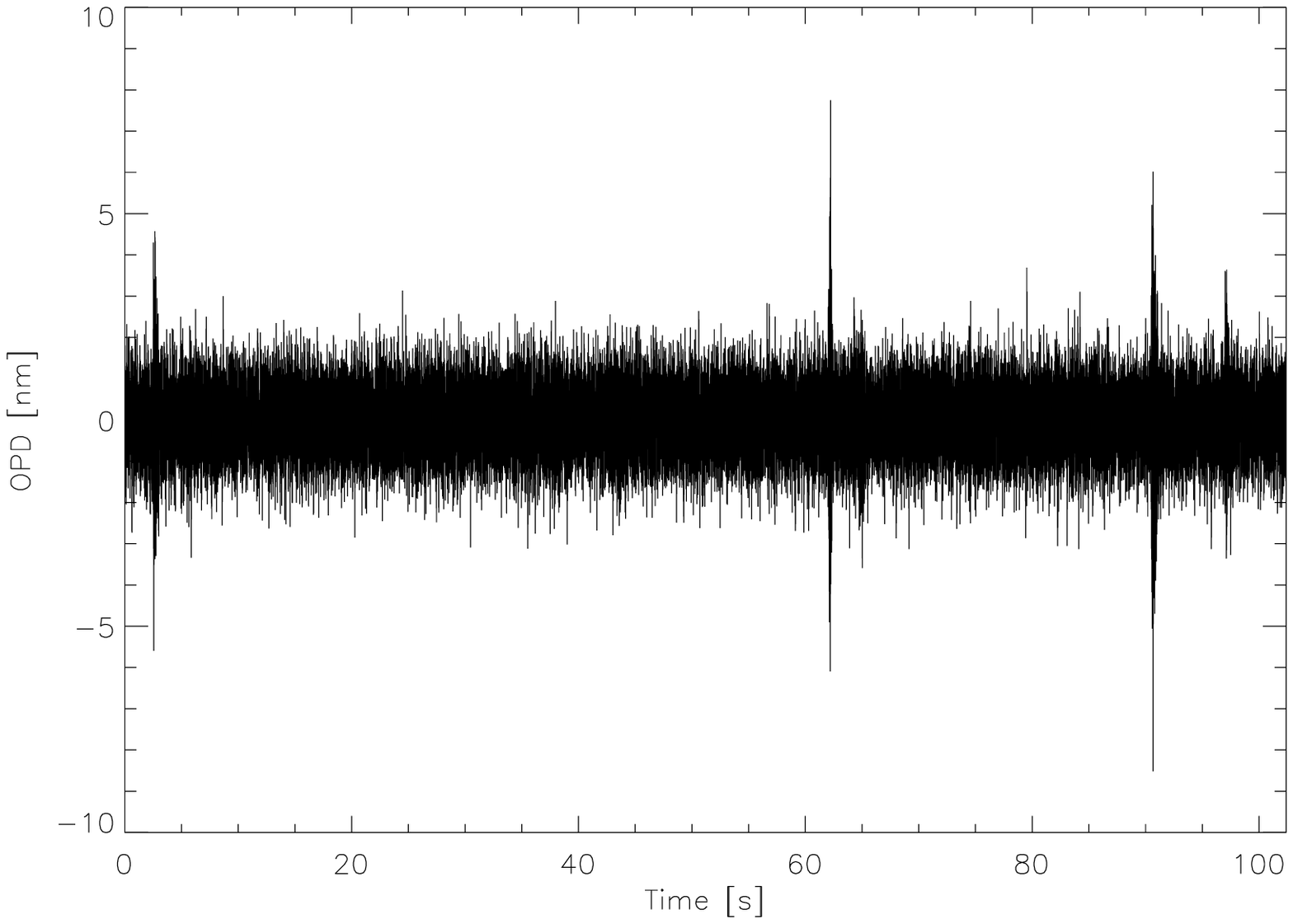}
    \end{tabular}
  \end{center}
  \caption{Simulation of the null rate (left) with the measured OPD (right).}
  \label{fig-opd}
\end{figure}

Figure~\ref{fig-opd} presents the contribution of the OPD on the mean null
rate. Compared to Fig.~\ref{fig-best-null}, we can see exactly the same
disturbances, with a similar amplitude. The OPD disturbances are characterized
by their standard deviation $\sigma_{OPD} = 0.80$~nm~rms. According to
Eq.~\ref{eq-system-nulling}, the contribution of the OPD on the null rate is
then

\begin{equation}
  N_\delta = \pi^2\moy[i]{\sigma_i^2}\moy[t]{(\delta-\delta_m)^2}, \text{ with }
  \delta_m = \frac{\moy[i]{\sigma_i^2\delta_i}}{\moy[i]{\sigma_i^2}}.
\end{equation}

That contribution can be reduced by having $\delta_{ref} = \delta_m$,
especially with the calibration described in Sec.~\ref{sec-opd-ref}. But for
this case, the calibration was not operationnal, so the search of the optimal
$\delta_{ref}$ was made manually.

So the contribution of OPD fluctuations and static error are

\begin{equation}
  N_\delta = \pi^2\moy[i]{\sigma_i^2}\left(\delta_{ref}^2+\sigma_{OPD}^2\right)
\end{equation}

In this case, we measured $|\delta_{ref}| = 0.48$~nm. So the OPD contribution
is $N_\delta = 1.7\times10^{-6}$. We can also calculate the contribution of
the null rate stability $\sigma_{N,\delta,T}$: we have
$\sigma_{N,T=2.58\mathrm{ms}}=1.8\times10^{-6}$ and
$\sigma_{N,T=1\mathrm{s}}=3.1\times10^{-7}$. Those values are almost equal to
the stability of the null rate, which shows that the null rate is only
disturbed by the OPD variations.

The reference position of the FS does not optimize the null rate of each
spectral channel, because of the phase dispersion
$\moy[t]{\Delta\phi_\lambda(t)}$. That dispersion is measured during the
correction of the OPD reference position presented in
Sec.~\ref{sec:CALIBRATION}.

The contribution of the chromatism to the null rate, is then~\cite{Serabyn01}
	
\begin{equation}
  N_{\lambda,i} = (\pi\sigma_i(\delta_i-\delta_{ref}))^2.
  \label{eq:null-lambda-i}
\end{equation}
	
So the total contribution on the whole spectral band is
	
\begin{equation}
  N_\lambda = \pi^2\moy[i]{(\sigma_i(\delta_i-\delta_{ref}))^2}.
  \label{eq:null-lambda}
\end{equation}

\begin{figure}
  \begin{center}
    \begin{tabular}{c}
      \includegraphics[width=0.48\linewidth]{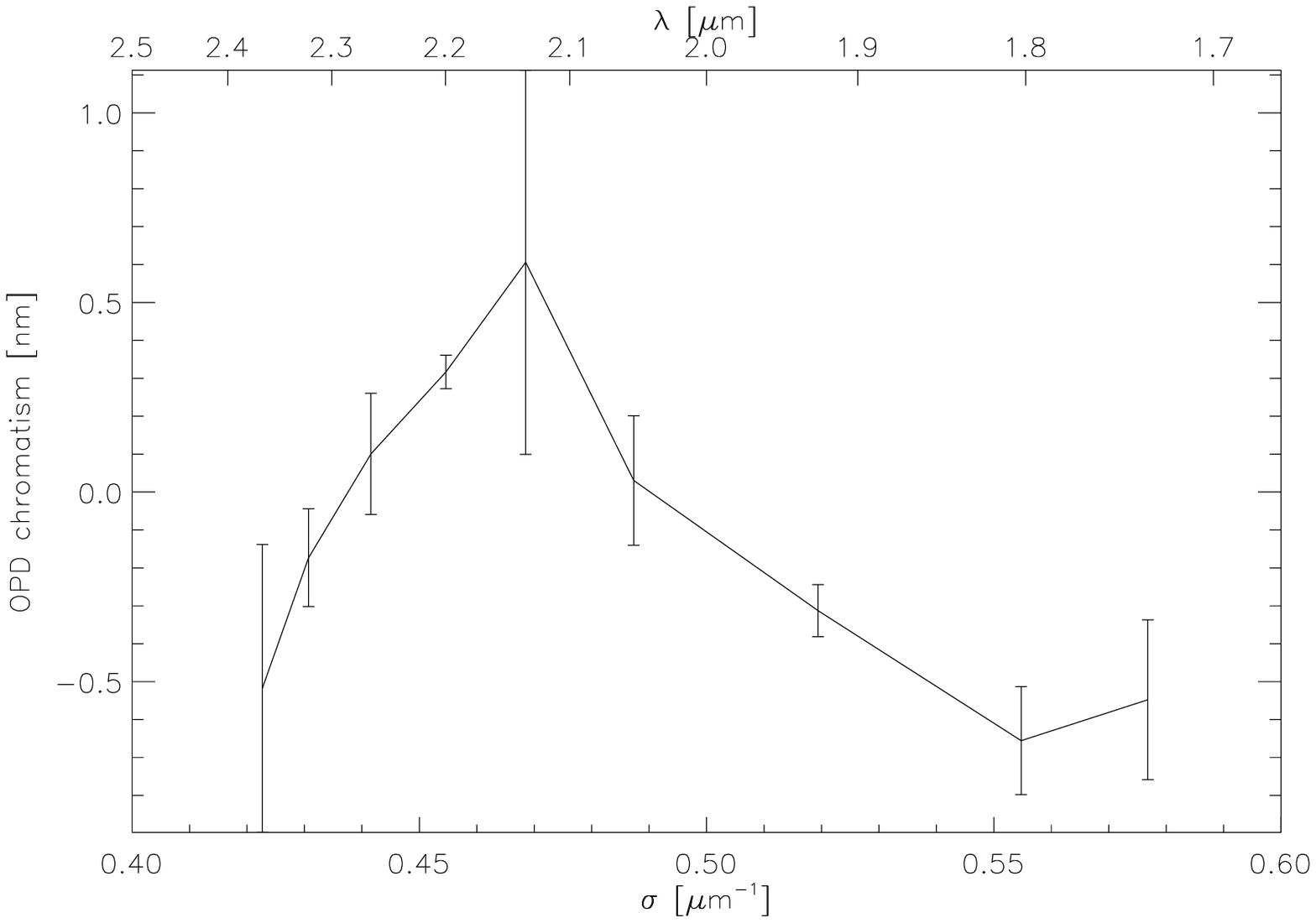}
      \includegraphics[width=0.48\linewidth]{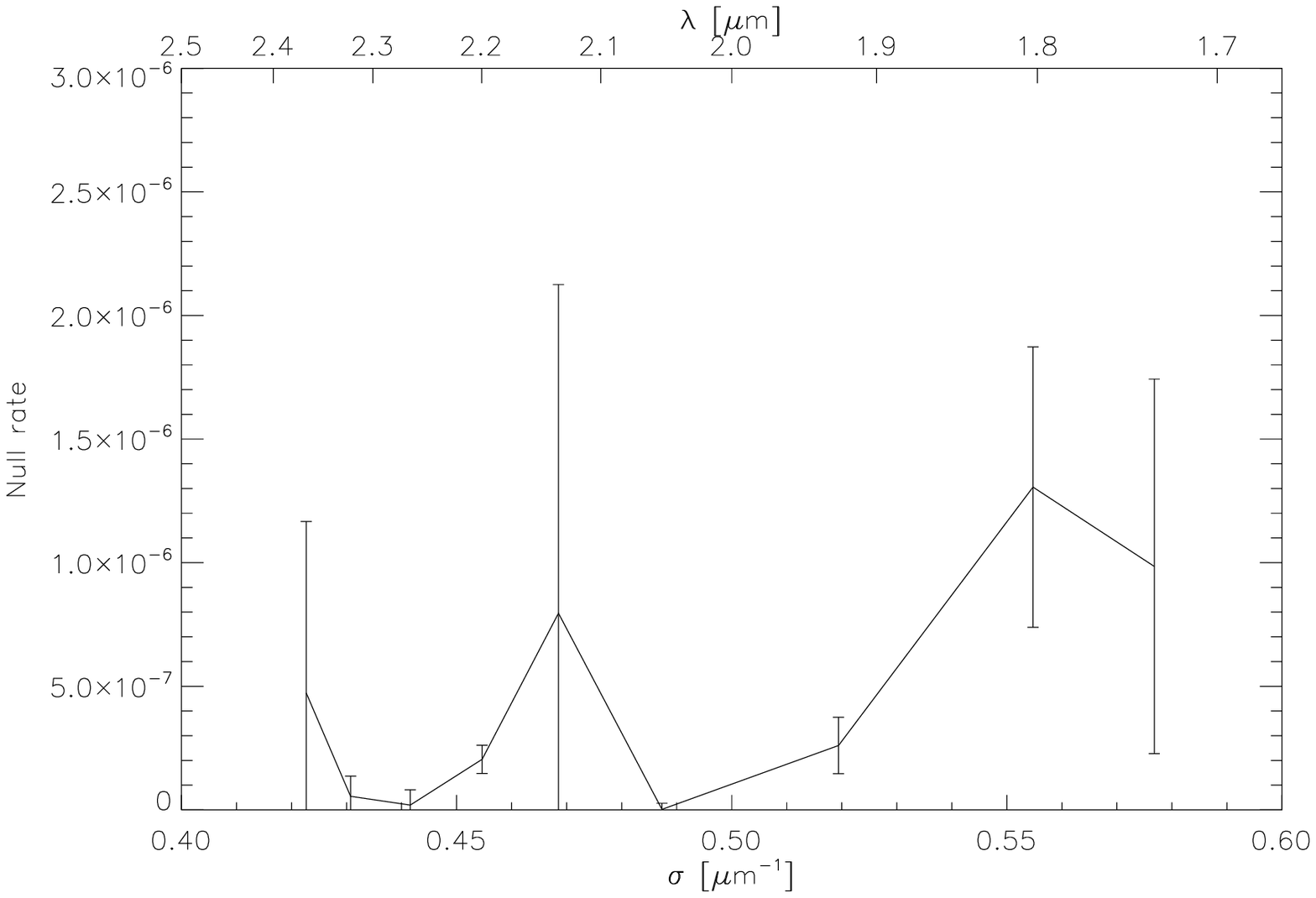}
    \end{tabular}
  \end{center}
  \caption{Chromatic dispersion of the positions of best null rates for the
    channels of the camera, and its contribution to the null rate.}
  \label{fig-chromatism}
\end{figure}

Figure~\ref{fig-chromatism} presents a measurement of the chromatic dispersion
on the channels of the camera. The contribution of the chromatism, calculated
with Eq.~(\ref{eq:null-lambda}), is then $N_\lambda = 4.6\times10^{-7}$. This
chromatism is due to the beam splitters of the MMZ, especially the thickness
differences of both plates and coatings, and also the difference in incidence
angle between the two arms.


\subsection{Stellar leakage}

The contributor corresponding to stellar leakage $N_*$ , corresponds in
Eq.~(\ref{eq-system-nulling}) to

\begin{equation}
  N_* = \frac{\pi^2}{16}(\theta_{dia}\sigma_{max}b)^2.
\end{equation}

The angular diameter $\theta_{dia}$ of the source is given by the ratio of the
diameter of the fiber $d$ and the focal length $f_0$ of the collimator. On
PERSEE, $d = 4.3$~\mum and $f_0 = 750$~mm, $b = 50$~mm and $\sigma_{max} =
0.61$~\imum, So the stellar leakage would be $N_* = 1.9\times10^{-2}$.

However, we obtained much better null rate. Indeed, this calculation applies
only for a spatially incoherent source, whereas in the case of PERSEE, the
source is spatially coherent, due to the single-mode fiber. Therefore, there
is no stellar leakage.


\subsection{Photometric mismatch}

The contribution of the photometric mismatch depends on the tip/tilt residue
of the control-loop, and the static photometric mismatch of the instrument.

From the Equation~(\ref{eq:flux-angle}), we can write

\begin{equation}
  \left\{
    \begin{array}{rcl}
      I_{a,i} &=& I_{a,i,max}(1-\frac{\displaystyle \pi^2}{\displaystyle
        5}\sigma_i^2D^2\zeta_a^2) \\[+6pt]
      I_{b,i} &=& I_{b,i,max}(1-\frac{\displaystyle \pi^2}{\displaystyle
        5}\sigma_i^2D^2\zeta_b^2) \\
    \end{array}
  \right.,
\end{equation}

with $\zeta_a^2 = (\alpha_a-\alpha_{a,ref})^2+(\beta_a-\beta_{a,ref})^2$ and
$\zeta_b^2 = (\alpha_b-\alpha_{b,ref})^2+(\beta_b-\beta_{b,ref})^2$.

Then the photometric mismatch $\varepsilon_i$ on each channel $i$ is

\begin{equation}
  \varepsilon_i =
  \varepsilon_{stat,i}+\frac{\pi^2}{10}\sigma_i^2D^2(\zeta_a^2-\zeta_b^2),
\end{equation}

with $\varepsilon_{stat,i} = \frac{\displaystyle
  I_{b,i,max}-I_{a,i,max}}{\displaystyle I_{b,i,max}+I_{a,i,max}}$. Then the
contribution of photometric mismatch on each channel is

\begin{equation}
  N_{\varepsilon,i} = \frac{\varepsilon_{stat,i}^2}{4}+
  \frac{\pi^4}{400}\sigma_i^4D^4(\zeta_a^2-\zeta_b^2)^2.
\end{equation}

The contribution on the null rate is then

\begin{equation}
  N_\varepsilon = \frac{\moy[i]{\varepsilon_{stat,i}^2}}{4}+
  \frac{\pi^4}{400}D^4\moy[i]{\sigma_i^4}(\zeta_a^2-\zeta_b^2)^2
  \label{eq:null-epsilon}
\end{equation}

\begin{figure}
  \begin{center}
    \begin{tabular}{c}
      \includegraphics[width=0.48\linewidth]{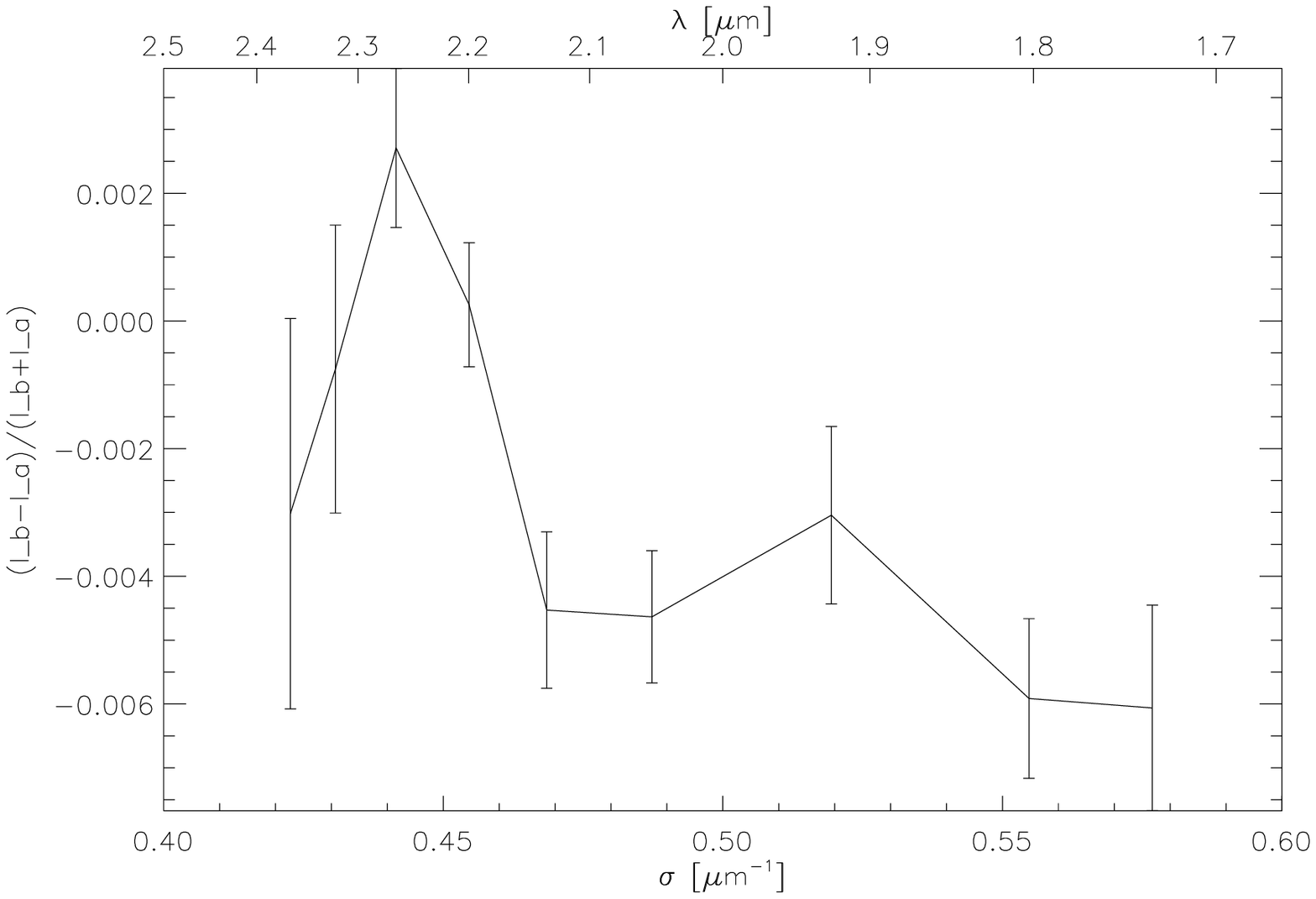}
      \includegraphics[width=0.48\linewidth]{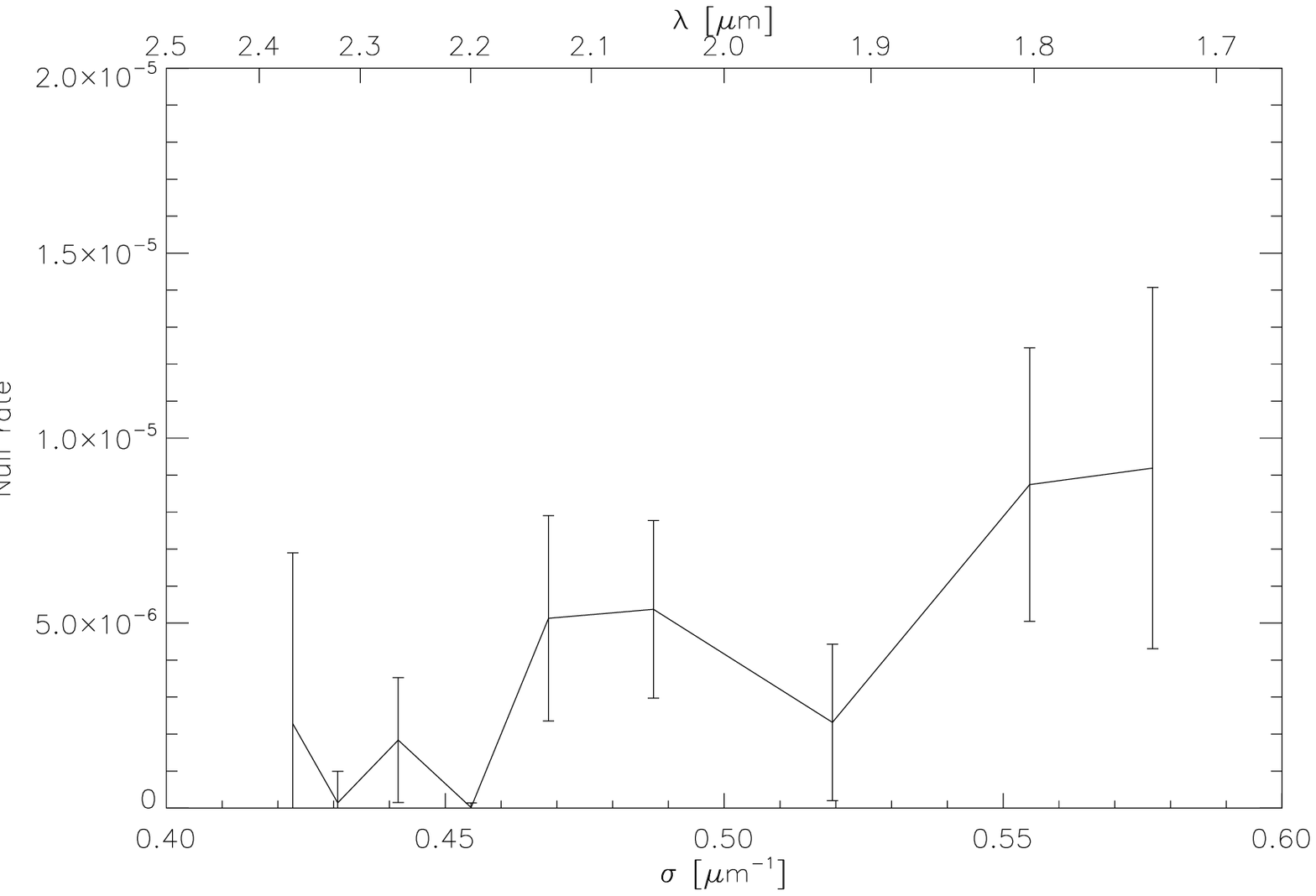}
    \end{tabular}
  \end{center}
  \caption{Photometric mismatch on the different channels of the camera.}
  \label{fig-epsilon}
\end{figure}

The static contribution of the photometric mismatch $\varepsilon_{stat,i}$ is
calculated during the calibration of the camera, and is only due to the
instrument optics. Figure~\ref{fig-epsilon} presents a measurement of
$\varepsilon_{stat}$, and its contribution on the null rate, calculated by

\begin{equation}
  N_{\varepsilon,stat,i} = \frac{\varepsilon_{stat,i}^2}{4}.
\end{equation}

Thus, the total contribution of instrumental photometric mismatch is given by

\begin{equation}
  N_{\varepsilon,stat} = \frac{\moy[i]{\varepsilon_{stat,i}^2}}{4}.
\end{equation}

That contribution is then $N_{\varepsilon,stat} = 3.9\times10^{-6}$.

The residue of tip/tilt control is also measured, and has a typical standard
deviation $\sigma_{tip/tilt,axis} = 56$~mas~rms on each axis, so the stability
of the position on each arm is $\sigma_{tip/tilt} = 80$~mas~rms (0.14\% of the
Airy disk).

According to Eq.~(\ref{eq:null-epsilon}), the dynamic contribution of the
control loop residue to the null rate is

\begin{equation}
  N_{\varepsilon,dyn} = \frac{\pi^4}{100}D^4\moy[i]{\sigma_i^4}\sigma_{tip/tilt}^4.
\end{equation}

Then, that contribution is $N_{\varepsilon,dyn} = 1.1\times10^{-11}$. It is
clearly negligible compared to $N_{\varepsilon,stat}$, but the relation in
$\sigma_{tip/tilt}^4$ is very sensitive to the FRAS correction residue.

For the stability over 100~s, the contribution of the tip/tilt control loop is
very faint, at $\sigma_{N,tip/tilt,T=1\mathrm{s}} = 1.4 \times 10^{-12}$, a
few decades under the contribution of the OPD control loop.


\subsection{Polarization study}

We have no way to measure the polarization contributions, but we can estimate
them from subtracting the other known contributions, by the equation:

\begin{equation}
  N_{pol} = N_m-N_\delta-N_\lambda-N_\varepsilon
\end{equation}

So, with $N_m = 8.9\times10{-6}$, $N_\delta = 1.7\times10^{-6}$, $N_\lambda =
4.6\times10^{-7}$ and $N_\varepsilon = N_{\varepsilon,stat} =
3.9\times10^{-6}$, we can estimate the contribution of polarization effects at
$N_{pol} = 2.8\times10^{-6}$.

So if that contribution is only attributed to the phase difference between the
polarizations $s$ and $p$, the OPD difference between the two polarizations is
$\delta_{s-p} = 2.1$~nm. On the contrary, if the contribution is only comming
from the angle of the polarizations of the two arms, that angle is
$\alpha_{rot} = 3.3$~mrad.


\subsection{Stability of the null rate over 7 hours}

We are currently conducting stability tests over a few hours (typically 10~h),
the typical duration of a target observation during a space mission. The first
test was made over 7~h during the night, with a sub-optimal alignment. The
average null rate is then higher than the value analyzed above.

\begin{figure}[b]
  \begin{center}
    \begin{tabular}{c}
      \includegraphics[width=0.48\linewidth]{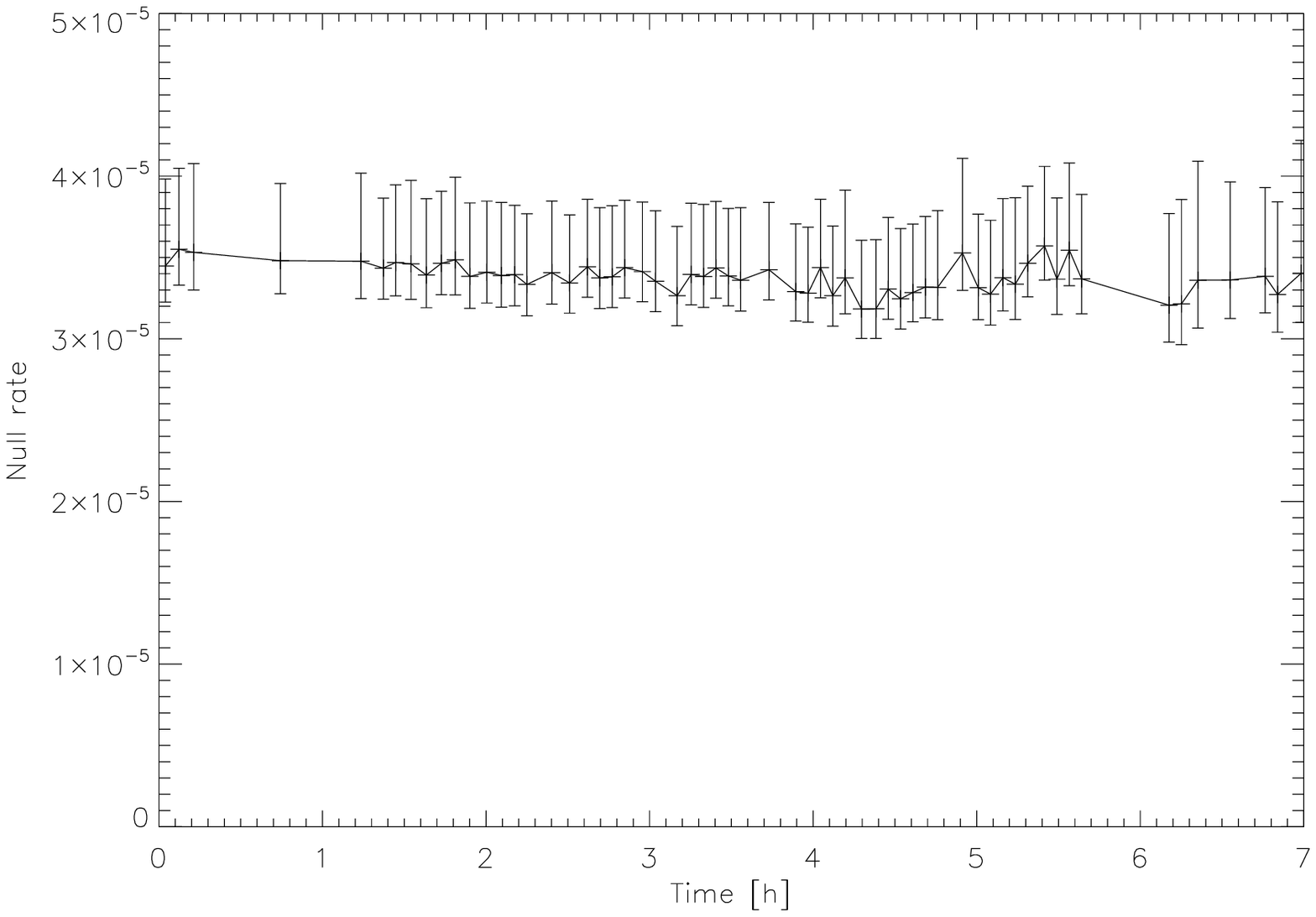}
      \includegraphics[width=0.48\linewidth]{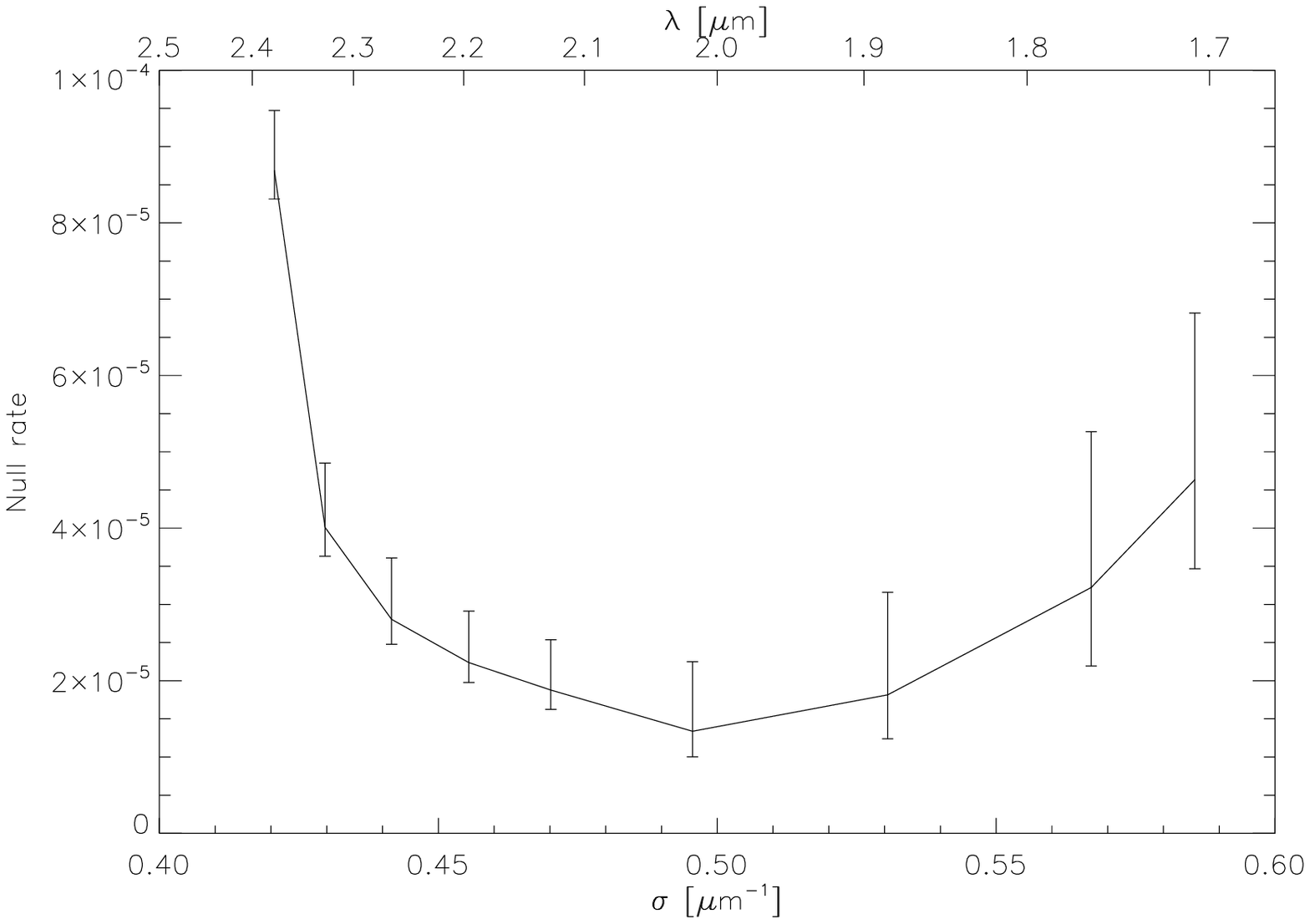}
    \end{tabular}
  \end{center}
  \caption{First measurement of null rate over 7~h: temporal sequence of the
    mean null rate (left) and temporal average of the null rates on the
    different spectral channels (right).}
  \label{fig-null-100s}
\end{figure}

Figure~\ref{fig-null-100s} presents the first long-term measurement of the
null rate over 100~s. We can see on the left figure that the null rate has a
higher value than the best result: $\moy[T=7\mathrm{h}]{N_m} = 3.37 \times
10^{-5}$, but shows a very good stability over 7~h:
$\sigma_{N,T=100\mathrm{s}} = 9 \times 10^{-7}$. That measured stability is a
decade better than the specification: $\sigma_{N,T=100\mathrm{s}} = 10^{-5}$.

In Fig.~\ref{fig-null-100s} (right), the null rate spectral profile looks like
the one presented in Fig.~\ref{fig-best-null}: the null rate is better at the
center of the spectrum than at the edges. The contribution of OPD and flux
mismatch are also higher at shorter wavelengths, and the contribution of
polarization should be higher at longer wavelengths. The chromatic dispersion
of the null rate is between $1.3 \times 10^{-5}$ and $8.7 \times 10^{-5}$, so
the amplitude is 7 times higher than on the best result.

\begin{table}
  \begin{center}
    \renewcommand{\arraystretch}{1.25}
    \begin{minipage}{10.5cm}
      \begin{tabular}{rccc}
        \multicolumn{4}{c}{OVERVIEW OF NULL RATE ANALYSIS}\\
        \hline Contribution & Specification & Best null (100~s) & Null rate
        (7~h) \\
        \hline $\moy[t]{N_{m}}$ & $10^{-4}$ & $8.9 \times 10^{-6}$ & $3.37 \times
        10^{-5}$ \\
        \hline $N_\delta$ & $3.5 \times 10^{-5}$ & $1.7 \times 10^{-6}$ & $2.8
        \times 10^{-6}$ \\
        $N_\lambda$ & $3.5 \times 10^{-5}$ & $4.6 \times 10^{-7}$ & $5.4
        \times 10^{-6}$ \\
        $N_\varepsilon$ & $2 \times 10^{-5}$ & $3.9 \times 10^{-6}$ & $3.3 \times
        10^{-6}$ \\
        $N_{pol}$\footnote{estimation based on the other contributions} &
        $10^{-5}$ & $2.8 \times 10^{-6}$ & $2.2 \times 10^{-5}$ \\
        \hline $\sigma_{N,T=1\mathrm{s}}$ & $1.5 \times 10^{-5}$ & $3 \times
        10^{-7}$ & $2 \times 10^{-7}$ \\
        \hline $\sigma_{N,\delta,T=1\mathrm{s}}$ & $1.5 \times 10^{-5}$ & $3
        \times 10^{-7}$ & $2 \times 10^{-7}$ \\
        $\sigma_{N,\varepsilon,T=1\mathrm{s}}$ & $3 \times 10^{-6}$ & $1.4
        \times 10^{-12}$ & $1.5 \times 10^{-11}$ \\
        \hline $\sigma_{N,T=100\mathrm{s}}$ & $10^{-5}$ & - & $9 \times 10^{-7}$ \\
      \end{tabular}
    \end{minipage}
    \caption{Overview of the analysis of the best null rate over 100~s and the
      null rate measured over 7~h, in average value and stability, and
      comparison with the original specifications of the bench.}
    \label{tab-null-100s}
  \end{center}
\end{table}

Table~\ref{tab-null-100s} summarizes the null rate analysis developped in this
section. In the near futur, we should be able to achieve an average null rate
of $10^{-5}$ over 10~h, with a stability better than $10^{-6}$, i.e. a decade
under the original specifications of the bench. As we also demonstrate the
capability of correcting injected disturbances, we should be able to achieve
that result even with typical disturbances on the bench.


\section{CONCLUSION}
\label{sec:CONCLUSION}

PERSEE showed a very high quality and performance well beyond the initial
specifications. The large majority of goals has already been demonstrated. We
have already presented some important lessons in Ref.~\citenum{Lozi10}, about
thermal stability and control loop feasibility. However, we can complete these
remarks by other conclusions.

Firstly, we elaborated a calibration scheme that can be applied to a space
mission, although it must take into account the linear-parabolic drift.That
procedure is fully automatic, and allows also to make an accurate analysis of
the instrument.

Secondly, we showed that a LQG controller can manage the disturbances of the
OPD, to achieve a very deep null rate. An experimental sub-nanometer control
has been demonstrated, even in the case of strong disturbances (more than
100~nm~rms reduced to 1~nm~rms). That result was very important for PERSEE,
because we work in the near infrared, which is more sensitive to OPD
variations than thermal infrared used in projects like DARWIN.

Finally, we showed that thanks to a a very efficient design, we obtained null
rates more than 10 times better than the initial conservative goal of
$10^{-4}$, thus close to the requirement of a nuller for the observation of
telluric planets. We achieved the deepest null ever achieved by a
near-infrared nulling interferometer: $8.9\times10^{-6}$ on a 37\% bandwidth,
between 1.65 and 2.45~\mum. PERSEE bench shows also a very good stability over
100~s, and also over 7~hours. We still have to confirm this deep null rate
over 10~hours, with the same type of stability, and with injected
disturbances.

The last steps will be the correction of linear-parabolic drifts, together
with the use of a new source module, which includes a planet and an
exozodiacal disk, presented in Ref.~\citenum{Henault11} during this
conference.


\acknowledgments

Julien Lozi's PhD is funded by CNES and Onera. PERSEE bench is supported by
CNES and the region \^Ile de France.



\end{document}